\documentclass{article}
\usepackage[english]{babel}
\usepackage{footmisc}
\usepackage{geometry}
\usepackage{lscape}
\usepackage{array}
\usepackage{amsmath}
\usepackage{amsfonts}
\usepackage{array}
\usepackage{graphics}
\usepackage{booktabs}
\usepackage{graphicx}
\usepackage{placeins}
\usepackage{subcaption}
\usepackage[labelfont=bf]{caption}
\usepackage[table,xcdraw]{xcolor}
\usepackage{amsmath}
\usepackage[utf8]{inputenc}
\usepackage[english]{babel}
\usepackage{natbib}
\usepackage{lscape}
\usepackage[backref=page,breaklinks=true,hidelinks]{hyperref}
\hypersetup{
    colorlinks,
    linkcolor={red!50!black},
    citecolor={blue!50!black},
    urlcolor={blue!80!black}
}
\renewcommand*{\backref}[1]{}
\renewcommand*{\backrefalt}[4]{%
    \ifcase #1 {\footnotesize(Not cited.)}%
    \or        {\footnotesize(Cited on page~#2.)}%
    \else      {\footnotesize(Cited on pages~#2.)}%
    \fi}

\usepackage{xfrac}
\usepackage{multirow}
\usepackage{bbm}
\usepackage{mathtools}
\usepackage{forest}
\usepackage{mdframed}
\usepackage{amsthm}
\usepackage{thmtools}
\usepackage{versions}
\usepackage{titlesec}
\usepackage{siunitx}

\newtheorem{application}{Application}

\newtheorem*{example*}{Example}

\newtheorem{proposition}{Proposition}
\newtheorem{assumption}{Assumption}

\newtheorem{definition}{Definition}
\newtheorem{corollary}{Corollary}
\newtheorem*{definition*}{Definition}

\newtheorem{theorem}{Theorem}

\makeatletter
\def\blfootnote{\xdef\@thefnmark{}\@footnotetext}
\makeatother
\usepackage{setspace}
\spacing{1.4}

\renewcommand{\thesection}{\Roman{section}}
\titleformat{\section}
  {\centering\Large\scshape}{\thesection.}{1em}{}

\renewcommand{\subsectionautorefname}{Section}

\DeclareMathOperator*{\argmin}{arg\,min}

\newcommand{\Ind}{\mathbbm{1}}  %
\renewcommand{\d}{\: \textnormal{d}}
\newcommand{\R}{\mathbb{R}}

\newcommand{\0}{\mathbf{0}}

\renewcommand{\P}{\mathop{}\!\textnormal{P}}
\newcommand{\E}{\mathop{}\!\textnormal{E}}

\newcommand{\X}{\mathcal{X}}
\newcommand{\Y}{\mathcal{Y}}

\newcommand{\F}{\mathcal{F}}  %
\newcommand{\ex}{\textsc{e}}  %
\newcommand{\Ex}{\mathcal{E}}  %

\newcommand{\const}{\text{const.}}

\usepackage{enumitem}

\addto\extrasenglish{%
}

\renewcommand{\subsectionautorefname}{Section}
\newcommand{\apxref}[1]{%
  \hyperref[#1]{Appendix~\ref*{#1}}%
}

\newcommand{\Srestrict}{S_1}
\newcommand{\Sexplain}{S_2}

\newcommand{\bigsize}{4}
\newcommand{\firstprob}{50}
\newcommand{\secondprob}{30}

\newcommand{\generalexplain}{
    \begin{tikzpicture}
        \path (0,{\bigsize*\secondprob/100}) rectangle ({\bigsize*(100-\firstprob)/100},\bigsize) node[pos=.5] {$\hat{f}(0,0)$};
        \path (0,0) rectangle ({\bigsize*(100-\firstprob)/100},{\bigsize*\secondprob/100}) node[pos=.5] {$\hat{f}(0,1)$};
        \path ({\bigsize*(100-\firstprob)/100},{\bigsize*\secondprob/100}) rectangle (\bigsize,\bigsize) node[pos=.5] {$\hat{f}(1,0)$};
        \path ({\bigsize*(100-\firstprob)/100},{\bigsize*\secondprob/100}) rectangle (\bigsize,0)  node[pos=.5] {$\hat{f}(1,1)$};

        \path (0,\bigsize) -- ({\bigsize*(100-\firstprob)/100},\bigsize) node[midway,above,scale=.7] {$x_1=0$};
        \path ({\bigsize*(100-\firstprob)/100},\bigsize) -- (\bigsize,\bigsize) node[midway,above,scale=.7] {$x_1=1$};
        \path (0,\bigsize) -- (0,{\bigsize*\secondprob/100}) node[midway,left,scale=.7] {$x_2=0$};
        \path (0,{\bigsize*\secondprob/100}) -- (0,0) node[midway,left,scale=.7] {$x_2=1$};
        \draw (0,0) rectangle (\bigsize,\bigsize);
        \draw ({\bigsize*(100-\firstprob)/100},0) -- ({\bigsize*(100-\firstprob)/100},\bigsize);     
        \draw (0,\bigsize*\secondprob/100) -- (\bigsize,\bigsize*\secondprob/100);         
    \end{tikzpicture}
}

\newcommand{\firstexplain}{
    \begin{tikzpicture}
        \path (0,0) rectangle ({\bigsize*(100-\firstprob)/100},\bigsize) node[pos=.5,align=center,scale=.7] {$(1{-}q)\: \hat{f}(0,0)$ \\ $+ \: q \: \hat{f}(0,1)$};
        \path ({\bigsize*(100-\firstprob)/100},0) rectangle (\bigsize,\bigsize) node[pos=.5,align=center,scale=.7] {$(1{-}q)\: \hat{f}(1,0)$ \\ $+ \: q \: \hat{f}(1,1)$};

        \path (0,\bigsize) -- ({\bigsize*(100-\firstprob)/100},\bigsize) node[midway,above,scale=.7] {$x_1=0$};
        \path ({\bigsize*(100-\firstprob)/100},\bigsize) -- (\bigsize,\bigsize) node[midway,above,scale=.7] {$x_1=1$};
        \path (0,\bigsize) -- (0,{\bigsize*\secondprob/100}) node[midway,left,scale=.7] {$x_2=0$};
        \path (0,{\bigsize*\secondprob/100}) -- (0,0) node[midway,left,scale=.7] {$x_2=1$};
        \draw (0,0) rectangle (\bigsize,\bigsize);
        \draw ({\bigsize*(100-\firstprob)/100},0) -- ({\bigsize*(100-\firstprob)/100},\bigsize);     
        \draw[dotted] (0,\bigsize*\secondprob/100) -- (\bigsize,\bigsize*\secondprob/100);         
    \end{tikzpicture}
}
    
\newcommand{\secondexplain}{

    \begin{tikzpicture}
        \path (0,{\bigsize*\secondprob/100}) rectangle (\bigsize,\bigsize) node[pos=.5,scale=.7] {$\sfrac{1}{2}\: \hat{f}(0,0) + \sfrac{1}{2} \: \hat{f}(1,0)$};
        \path (0,0) rectangle (\bigsize,{\bigsize*\secondprob/100}) node[pos=.5,scale=.7] {$\sfrac{1}{2}\: \hat{f}(0,1) + \sfrac{1}{2} \: \hat{f}(1,1)$};

        \path (0,\bigsize) -- ({\bigsize*(100-\firstprob)/100},\bigsize) node[midway,above,scale=.7] {$x_1=0$};
        \path ({\bigsize*(100-\firstprob)/100},\bigsize) -- (\bigsize,\bigsize) node[midway,above,scale=.7] {$x_1=1$};
        \path (0,\bigsize) -- (0,{\bigsize*\secondprob/100}) node[midway,left,scale=.7] {$x_2=0$};
        \path (0,{\bigsize*\secondprob/100}) -- (0,0) node[midway,left,scale=.7] {$x_2=1$};
        \draw (0,0) rectangle (\bigsize,\bigsize);
        \draw[dotted] ({\bigsize*(100-\firstprob)/100},0) -- ({\bigsize*(100-\firstprob)/100},\bigsize);     
        \draw (0,\bigsize*\secondprob/100) -- (\bigsize,\bigsize*\secondprob/100);         
    \end{tikzpicture}
}

\title{
Unpacking the Black Box: \\ Regulating Algorithmic Decisions}
\author{Laura Blattner 
\and Scott Nelson
\and Jann Spiess }

\date{May 2024}

\begin{document}
\renewcommand{\subsectionautorefname}{Section}
\renewcommand{\footnoteautorefname}{Footnote}

\maketitle

\blfootnote{%
\hspace{-2.5em}
Authors are listed in alphabetical order.

We thank Mario Curiki, Georgy Kalashnov,
 and Ruying Gao for outstanding research assistance.  We thank Susan Athey, 
Simon Freyaldenhoven, Talia Gillis, Paul Goldsmith-Pinkham, Damian Kozbur, Danielle Li, Sendhil Mullainathan, Ashesh Rambachan, Amit Seru, Ken Singleton, PR Stark, Chenzi Xu, Louis Kaplow, Kathryn Spier, the FinRegLab team, and seminar and conference participants at Stanford, Yale, Harvard, Zurich, Oxford,
the NBER Summer Institute,
Stanford SITE,
the New Perspectives on Consumer Behavior in Credit and Payments Markets Conference,
the AEA Annual Meeting,
the FTC, 
ENSAI, 
the OCC,
EC,
the Artificial Intelligence and Big Data in Finance Research (ABFR) Forum,
NASMES,
the Machine Learning in Economics Summer Institute,
and
the 2nd Zurich AI + Economics Workshop
for helpful discussions and comments. We thank the Stanford Institute for Human-Centered Artificial Intelligence (HAI) and Amazon Web Services for generous support. Any errors or omissions are the responsibility of the authors.

This manuscript supersedes an earlier version that was accepted and presented at EC'22, with an extended abstract published as:
Blattner, Laura, Scott Nelson, and Jann Spiess (2022). Unpacking the Black Box: Regulating Algorithmic Decisions. In \emph{Proceedings of the 23rd ACM Conference on Economics and Computation (EC'22)}, page 559.

}

\begin{abstract}
\noindent
What should regulators of complex algorithms regulate?
We propose a model of oversight over `black-box' algorithms used in high-stakes applications such as lending, medical testing, or hiring.
In our model,
a regulator is limited in how much she can learn about a black-box model deployed by an agent with misaligned preferences.
The regulator faces two choices: first, whether to allow for the use of complex algorithms; and second, which key properties of algorithms to regulate.
We show that limiting agents to algorithms that are simple enough to be fully transparent is inefficient as long as the misalignment is limited and complex algorithms have sufficiently better performance than simple ones.
Allowing for complex algorithms can improve welfare, but the gains depend on how the regulator regulates them.
Regulation that focuses on the overall average behavior of algorithms, for example based on standard explainer tools, will generally be inefficient.
Targeted regulation that focuses on the source of incentive misalignment, e.g., excess false positives or racial disparities, can provide second-best solutions.
We provide empirical support for our theoretical findings using an application in consumer lending,
where we document that complex models regulated based on context-specific explanation tools outperform simple, fully transparent models.
This gain from complex models represents a Pareto improvement across our empirical applications that is preferred both by the lender and from the perspective of the financial regulator.
\end{abstract}

\clearpage

\section{Introduction}

The increasing adoption of complex prediction algorithms in sensitive applications such as hiring, lending, medical testing, college admissions, and pre-trial detention raises the questions how a regulator should regulate them.
On the one hand, when algorithms replace human decision-makers, a regulator can now analyze and intervene in decision processes more systematically.
On the other hand, the increasing complexity of artifical-intelligence algorithms makes this goal elusive, and regulation often has to rely on an imperfect understanding of complex black-box algorithmic decisions.
These concerns have led to calls for restricting the complexity of algorithms and relying on simpler, fully transparent decision rules.

This article captures trade-offs between complexity and oversight of algorithms in a principal--agent model with misaligned preferences.
We consider a delegation game between a principal who regulates a machine-learning algorithm and an agent who has the technology to build it.
There is an incentive conflict between agent and principal, but the principal is limited in how much she can learn about the agent's black-box prediction model, and she has to regulate based on limited information about the model instead.
Within this framework, we ask what a regulator of algorithms should regulate.
We show that a restriction to fully transparent, simple algorithms that fully aligns choices comes at a large cost.
As an alternative, we consider the regulation of complex models based on simple model explanations, and show that appropriately designed explanations that capture the misalignment between principal and agent provide a second-best solution.
We then show the applicability of these results in an empirical application to consumer lending in a large credit bureau dataset.

Decision-makers increasingly rely on complex prediction algorithms to make high-stakes decisions.
In many such settings, incentive conflicts arise between those building the prediction tools and the entities tasked with overseeing their use.
An employer might worry about a hiring manager using a prediction model that produces low job offer rates for minority job applicants.
A financial regulator might worry about lenders' risk models under-predicting credit risk to enable increased leverage.
An insurance company might worry about a hospital's prediction model over-predicting the risk of heart attack leading to costly over-testing.
A key challenge for algorithmic oversight is to determine when such complex algorithmic prediction functions represent the principal's preferred choices, and when they reflect misaligned incentives.

We capture the incentive conflicts between an agent who designs a prediction algorithm and a principal who regulates its use in a principal--agent model.
Our setup mirrors that of a classical delegation problem, where the agent chooses a prediction function according to his preferences, but is subject to constraints set by the principal.
For example, a company may restrict which algorithms can be used for hiring, a financial regulator may put constraints on risk-scoring models used by banks, and an insurance company may set standards for the screening algorithms used in a hospital.
Classical delegation models going back to \cite{holmstrom1977} and further developed in e.g. \cite{alonso_optimal_2008} and \cite{Frankel2014-yn} suggest solutions that partially restrict agent choices in a way that increases alignment, while still leaving enough flexibility to leverage the agent's technology and private information.

In this article, we depart from standard delegation models by assuming that the agent's choice of prediction function may be too complex to be fully available to the principal.
Specifically, we assume that the prediction function chosen by the agent comes from a black-box machine learning model with so many parameters that fully communicating it may not be feasible.
This restriction is motivated by the increasing use of very complex AI models in high-stakes decisions, such as deep neural networks and boosted tree models.
The constraint may also be motivated by concerns around releasing proprietary algorithms that the maker of the algorithm may want to protect, precluding the communication of the full prediction function to the principal.

Instead of accessing the full algorithm, we assume that the principal can reglate an algorithm only based on a simple description of the complex prediction function.
Building upon work in computer science, we call this simple description an ``explanation'' of the complex model.
For example, the behavior of a complex neural network may be described in terms of a few key variables that explain a good fraction of the model's behavior, such as those identified by tools like SHAP \citep{lundberg2017unified}.
Similarly, a complex tree model may be described in terms of those covariates deemed most important according to their contribution to splitting the data.
In our formal model, we capture this idea by assuming that the principal can only capture a simple (linear) projection of the complex prediction model on a few key variables.
Consequently, the principal misses some of the behavior of the complex model.

A first response to the limited ability of the principal to capture complex models could be to restrict the agent to fully transparent, simple models that can be fully regulated.
Within our framework, we argue that restrictions to fully explainable models come at a potentially large cost.
For example, an employer could require a hiring manager to rely on simple decision trees,
a financial regulator may require a bank to score creditworthiness only using a relatively interpretable logistic regression, 
or an insurance company may require a hospital to use a transparent decision rule for assessing the risk of heart attack. 
While fully aligning choices, such restrictions also reduce the efficiency of the prediction functions.
As a result, we show formally that simple, fully transparent algorithms can only be optimal when the principal's first-best solution is easier to describe in simple terms than is the difference between the agent's and the principal's preferred prediction functions.
In the above examples, this would usually require that the misalignment in preferences is very large, or that the loss from simple algorithms is small.

As an alternative, we consider second-best regulation that allows for complex predictions and only regulates based on their simple explanations.
For example, a lender may still be allowed to use fully complex deep-learning algorithms to predict repayment, but has to choose among models for which a simple explanation indicates alignment with the regulator's risk preferences.
While the principal will not be able to align preference optimally in this case, this solution can provide a more efficient trade-off between efficiency and alignment than simple algorithms, unless the misalignment is large or hard to detect by the explainer.

We also show that it matters which specific model explanations regulation is based on.
Regulation based in an explaination that focuses on preserving the most information about the average behavior of the prediction function -- we term this the ``agnostic explainer'' -- is generally inefficient.
This approach is the focus of many available explainer tools, and we argue that we can improve over it in our context.
Instead, optimal regulation regulates based on features related to preference misalignment. Intuitively, this ``targeted explainer'' inspects parts of the algorithm function that are most likely to reflect the preference misalignment.
This targeted regulation can achieve high utility for the principal as long as the difference in agent and principal goals can be captured well by some low-dimensional representation.
Improved explainer tools for the regulation of algorithms in critical applications should therefore be specific to the context and nature of preference misalignment.
For example, if our concern is unfair hiring decisions, then regulation should capture features that are particularly related to differences across protected groups.
If group membership itself is observable, then the optimal explainer in this case focuses on group membership itself as in \cite{kleinberg2018algorithmic}, but our characterization extends to cases when only correlates of membership are observable. 
If we are concerned with a bank taking on loans that are especially high-risk in a stress-test scenario, then our description of complex credit scores should focus on features that are particularly likely to identify risk in that scenario. 
And an insurance company may want to audit a hospital's prediction model specifically based on features that correlate with over-testing.

We apply our framework for overseeing complex algorithms to the regulation of unsecured consumer lending in an empirical case study using large-scale credit bureau data.
We consider a lender who builds a credit-scoring function using boosted trees from over 500 features on a training dataset of over 200,000 borrowers.
We then consider two types of regulators overseeing the lender.
A first regulator is concerned with the fairness of credit scores, and has a preference for low disparate impact across protected groups.
A second regulator has different risk preferences from the lender, and wants the lender to deploy a credit score that predicts default well in a bad state of the economy.
In both cases, the regulator has to rely on simple descriptions of the complex credit score in terms of a few key variables whose role is measured by a linear regression.
We then compare different policy options and explainers on a hold-out dataset.

Across both applications in our empirical exercise, we document that well-regulated complex models improve over simple, fully transparent solutions.
Restricting the lender to a simple linear regression with a few key variables leads to suboptimal fit and moderate disparate impact in the first application.
Here, moving to complex models with simple explainer constraints can improve fit, while also reducing disparities across groups.
Similarly, complex models can improve efficiency across all states of the economy in the risk application.
Our empirical results therefore provide an example for our theoretical finding that the cost of full transparency may be too large when effective alternatives in the form of simple model descriptions are available.

Our empirical example also documents the value of targeted regulation over agnostic explainers. While either approach offers a Pareto improvement over restrictions to very simple models, the best outcomes for both regulator and lender are achieved with a targeted explainer, which focuses on those aspects of the credit score that are particularly relevant for aligning preferences, rather than with an agnostic explainer, which focuses on those covariates that best summarize the credit scoring function on average. 
In the case of disparate impact, the targeted explainer focuses particularly on covariates that are most related to differences across groups.
With different risk preferences, the targeted explainer considers those variables that are indicative of changes in repayment behavior across states of the economy.
The benefit of improved regulation is best illustrated by its impact on disparate impact in approval rates.
While an unconstrained lender would be 4.5 percentage points more likely to approve non-minority applicants, this difference is reduced to 0.9 percentage points with an agnostic explainer, and to close to zero when the explainer is targeted to disparate impact.
Despite the explanation constraint, the performance of the credit score as a risk predictor remains good.
This stands in contrast with restricting the lender to a very simple model, which has considerably worse prediction fit while reducing disparate impact by only 1.0 percentage points.

\paragraph{Related literature and contributions.}

Our work contributes to a nascent literature that studies algorithmic decision-making \citep[e.g.][]{athey_allocation_2020} and how to regulate it.  Most work in this area has focused on trade-offs between algorithmic fairness and performance: \citet{Gillis2019-pc}, \citet{coston2021characterizing}, and \citet{gillis2021input} 
examine the design and limits of algorithmic fairness audits; \cite{dwork2012fairness}, \citet{corbett2017algorithmic}, \citet{corbett2018measure}, and \citet{yang2020equal} consider ex-ante (i.e., pre-deployment) fairness constraints for algorithms; \cite{liang2023algorithm} characterize how varying the information available to an algorithm can trade off between fairness and model performance. These trade-offs are sometimes understood as tracing out a fairness-vs.-accuracy Pareto frontier \citep{menon2018cost, little2022fairness, liang2023algorithm, meursault2022one}, such that the role of social planner can be seen as choosing a preferred outcome on the frontier \citep{kearns2019ethical}.\footnote{See also critical discussions of existing approaches to fairness in \cite{kasy2021fairness} and \cite{kasy2023algorithmic}, and the broader discussion of whether algorithms and artificial intelligence are aligned with societal goals in \citet{korinek2022aligned}.} Most related to our approach, \cite{rambachan_economic_2020} study the regulation of algorithmic fairness in a principal-agent framework, where a social planner (principal) may not be able to achieve outcomes on a first-best frontier, and where second-best feasibility becomes relevant. Related concerns about the ability of a planner to interpret an algorithm have prompted calls for lower model complexity \citep{rudin2019stop}, debates about model interprability \citep{doshi-velez_towards_2017,lipton2018mythos}, and analyses of optimal algorithmic transparency \citep{sun2021algorithmic}. 

A parallel literature examines limits to the efficiency of algorithmic decisions, for example due to algorithmic bias  \citep[e.g.][]{lambrecht2019algorithmic, arnold2022measuring, fuster2022predictably}. This work has asked whether such biases are worse than their analogs from simpler or non-algorithmic decisionmaking \citep[e.g.][]{cowgill2017algorithmic, agrawal2019exploring, chan2022selection, arnold2022measuring}, whether algorithms necessarily inherit biases from human-generated training data \citep[e.g.][]{rambachan2019bias,cowgill2019economics,barocas2016big,angelova2023algorithmic}, and whether algorithms are able to learn over time about disadvantaged groups that the algorithm initially evaluates poorly \citep{li2020hiring}. While often not expressed in a principal-agent framework, these concerns about algorithmic bias closely parallel the agency conflicts we study: an algorithm designer concerned about bias in her algorithm can use the same regulatory approach we provide, here viewing her algorithm as her agent.

Relative to these literatures, we make three contributions. First, we offer a framework that nests many types of potential incentive misalignment between a developer of an algorithm and social planner, or between any algorithmic agent and a principal: these include a broad set of distributional objectives and fairness concerns, as well as, for example, diverging risk preferences. Second, existing analyses of algorithmic audits and other algorithmic regulation often assume that disclosure of all underlying algorithmic inputs (data, training procedure, and decision rule) is possible. We study a world in which regulators will have access only to parts of this information, for example a simplified representation of the credit scoring model. Given the complexity of machine learning and artificial intelligence tools, and potential limitations on the technical or legal reach of regulators, it is important to study optimal algorithmic regulation under informational constraints. Third, we provide empirical validation for our theoretical results in a real-world, economically important setting, where we examine the regulation of consumer credit scoring. As in our Theorem \ref{thm:optimalregulation}, this empirical application shows that both a regulator and a lender can achieve privately preferred outcomes when the lender is permitted to use complex algorithmic credit scoring, rather than being constrained to simple and explainable models. 

From an empirical perspective, we also contribute to a body of work on the role of default prediction models in US consumer finance. Most of this work explores properties of these models and their benefits, for example through overcoming adverse selection \citep{einav_impact_2013, adams_liquidity_2009}, deterring moral hazard \citep{chatterjee_quantitative_2020}, and facilitating loan securitization \citep{keys_lender_2012, keys_did_2010}. Related work also warns that algorithmic underwriting can shape disparities in credit misallocation \citep{blattner_laura_how_2021}, reduce loan approval rates for disadvantaged groups \citep{fuster2022predictably}, and perpetuate cross-group disparities in loan terms \citep{bartlett_algorithmic_2019}. These concerns motivate our work to study optimal algorithmic regulation and highlight some of the sources of preference misalignment we study in our theoretical framework.

We differ from the literature on optimal \emph{public} disclosure in the financial system \citep[e.g.][]{goldstein_stress_2017,faria-e-castro_runs_2017,williams_stress_2017,judge_stress_2020}, and the literature on firm disclosure more broadly \citep[e.g.][]{leuz_economic_2000,greenstone_mandated_2006}, as we focus on private disclosure to a regulator when decisions are automated and based on complex risk prediction algorithms.  We assume there are (technical, practical, or legal) limitations on the amount of information the regulator can obtain about the algorithms and ask what optimal regulation looks like given these constraints. This approach differs from the existing literature which assumes that regulators can exercise choice over how much information to request.

More broadly, we add to a growing literature in computer science that studies algorithmic audits and derives specific explainability techniques from axioms about their deployment-agnostic properties \citep[e.g.][]{bhatt_explainable_2020,carvalho_machine_2019,chen_interpretable_2018,doshi-velez_towards_2017,guidotti_survey_2018,hashemi_permuteattack_2020,lundberg2017unified,murdoch_definitions_2019,ribeiro_why_2016}.  In particular, \cite{lakkaraju_how_2020}, \cite{slack_fooling_2020}, and \cite{lakkaraju_faithful_2019} study the limitations of post-hoc explanation tools in providing useful and accurate descriptions of the underlying models, and show that simple explanations can be inadequate in distinguishing relevant model behavior.
Relative to these contributions, we show that the optimal regulatory design for algorithms with partial information depends on the nature of preference misalignment that motivates regulation. In other words, we highlight that explaining or interpreting a model inherently requires an understanding of the objectives of that explanation or interpretation,
while purely technical or axiomatic approaches may miss important welfare-relevant consequences of model behavior.
We also highlight some limitations of recent debates around the interpretability and explainability of prediction models.
Embracing our utility optimization framework, we show that requiring a model to be fully explainable or interpretable can be misguided since it may force an agent to sacrifice model flexibility in ways that reduce rather than increase welfare.

In the long-standing literature on delegation under moral hazard, our setup can be viewed as a multi-tasking problem where each dimension of the agent’s scoring rule is a separate hidden action. In contrast with the literature’s traditional focus on how to use heterogeneously noisy signals of these hidden actions in an incentive scheme \citep{holmstrom1991multitask,holmstrom1994firm}, we study the principal’s choice over which dimensions of these hidden actions -- or which low-dimensional representation of them -- she wants to observe in an audit. One closely related finding to ours is \cite{baker1992incentive}, which studies optimal incentive schemes under incentive misalignment over multiple actions, though \cite{baker1992incentive} takes the information structure as given rather than chosen by the principal. Alternatively, our setting can also be viewed as a delegation problem \citep{holmstrom1977} that considers constaints on the set of actions allowed to the agent \citep{melumad1991communication,alonso_optimal_2008,Frankel2014-yn}. Related work on principals designing the information structure is also found in the Bayesian persuasion literature \citep{kamenica_bayesian_2011}, though our setting is one of monitoring by the principal rather than of persuasion.

\paragraph{Structure of this article.}

The remaining article is organized as follows:
\autoref{sec:model} sets up our model and presents the main theoretical results. \autoref{sec:empirical} lays out our empirical implementation. \autoref{sec:conclusion} concludes.

\section{A Model of Regulation with Complexity Constraints}
\label{sec:model}

In this section, we model the regulation of algorithms as a two-player game between a principal and an agent.
The principal delegates the choice of a prediction function to the agent.
The agent receives a training signal and chooses the prediction function subject to constraints set by the principal.
The principal can not fully observe the potentially complex prediction function chosen by the agent,
and instead has to rely on a simple description.
The preferences of agent and principal are not necessarily aligned.

Based on this model, we consider trade-offs between complexity and oversight.
We study two types of policies and show how they affect equilibrium outcomes.
First, we consider the case where there are ex-ante restrictions on algorithms, which force the agent to use only simple prediction functions that the principal can understand completely.
Alternatively, we consider the case in which the agent chooses among complex algorithms and the principal restricts the agent's choice based on simple descriptions only.
We characterize the trade-off between these two policy tools -- ex-ante restrictions vs. ex-post explanations.
We then describe the optimal design of simple ex-post explanations of complex prediction functions.

\subsection{Model setup}
\label{subsec:modelsetup}

A principal oversees the choice by an agent of prediction function $f \in \F$.
The agent receives a training signal $\theta \in \Theta$, and chooses a prediction function $f$ to maximize utility $U^A_\theta(f)$.
The principal can constrain the choice of prediction function $f$, and has a preference over prediction functions expressed by utility $U^P_\theta(f)$.

For concreteness, we assume that the prediction function represents a mapping $f: \X \to \R$ and that the training signal $\theta$ parametrizes a distribution $\P_\theta$ that implies a joint distribution over a response variable $y \in \Y$ and a covariate vector $x \in \X$.
Throughout, we assume that the agent's preference can be expressed as an average gain $U^A_\theta(f) = \E_\theta[u(f(x),y)]$ from deploying the function $f$ on data $(y,x)$ that follows the distribution $\P_\theta$.
As examples of the utility function $u(f(x),y)$, for continuous $y$ the agent may care about minimizing the mean-squared prediction error of $f(x)$,  $u(f(x),y) = - (y - f(x))^2$.
Or the agent may want to maximize profit $u(f(x),y) = \Ind(f(x) \geq \bar{f}) (r \: y - c \: (1-y))$ from classifying instances $x$ with binary outcome $y \in \{1,0\}$ correctly according to a threshold $\bar{f}$, where $r$ is the return for a correctly classified $y=1$ instance and $c$ the cost of misclassifying a $y=0$ instance.
As we lay out in the following applications, such a prediction $f(x)$ can correspond to a score that measures the aptitude of job applicants and the classification $\Ind(f(x) \geq \bar{f})$ to the decision whether to invite the applicant for an interview, where $r$ is the return to a successful interview and $c$ the cost of an unsuccessful one.
Similarly, $f(x)$ can be a credit score that estimates the repayment probability of loan applicants used to determine who gets a loan,
or a predicted incidence of a disease used to make testing decisions.
In these cases, $u(f(x),y)$ would be the gain of the agent in classifying an individual instance with true response $y$ as $f(x)$, and $U^A_\theta(f)$ the overall benefit across instances.
The principal's preference $U^P_\theta(f)$ may, for example, differ from the maximization of expected utility by distributional considerations, different risk preferences, or an emphasis on different target populations.
In order to accommodate such cases, we assume that $\P_\theta$ may imply a distribution over additional random variables or the membership in a subgroup of the population.

\begin{application}[Disparate impact in job screening]
    \label{app:DI}
    A company delegates the screening of applicants to a manager. The manager aims to score applicants in a way that is most reflective of future work performance, while the company also cares about ensuring that the screening process does not discriminate against protected groups and that interviewers get to see a diverse group of applicants. Specifically, we assume that the manager chooses scores $f(x)$ to maximize expected utility $U^A_\theta(f) = \E_\theta[u(f(x),y)]$, while the company attaches a penalty to average differences between male and female applicants, $U^P_\theta(f) = \E_\theta[u(f(x),y)] - \lambda (\E_\theta[f(x)|g{=}1] - \E_\theta[f(x)|g{=}0])$, where $g$ is a binary indicator for group membership and $g=1$ denotes male applicants.%
    \footnote{%
        \label{fn:DI}%
        Disparate impact, as expressed here, is only one of multiple ways in which we could capture a concern about discrimination \citep[e.g.][]{kleinberg2016inherent,chouldechova2017fair}.
        Even for disparate impact, we could alternatively consider measures of conditional parity, use a squared penalty $\lambda (\E_\theta[f(x)|g{=}1] - \E_\theta[f(x)|g{=}0])^2$ to express a concern with larger average differences, and/or formulate disparities in terms of the implied classification decision.
        We opt for a simple linear formulation that allows seamless integration into our later theory, which can be understood as the Lagrange version of a restriction on $\E_\theta[f(x)|g{=}1] - \E_\theta[f(x)|g{=}0] \leq c$ that expresses a concern with the scoring of the protected group.}
\end{application}
            
\begin{application}[Varying risk preferences in credit provision]
    \label{app:risk}
    A financial regulator oversees the credit provision of a lender. The lender aims to maximize overall expected utility $U^A_\theta(f) = \E_\theta[u(f(x),y)]$ from credit scoring, while the regulator cares about the potential downside in a bad (or ``low'') state of the economy, $U^P_\theta(f) = \E_\theta[u(f(x),y)|s{=}\text{low}]$. In that rare low state, the regulator believes that the probability $\E_\theta[y|x,s{=}\text{low}]$ of repayment is lower than their repayment probability $\E_\theta[y|x,s{=}\text{high}]$ in the (more common) high state, leading to excess risk taken on by the lender.
    Here, we assume that the future state $s$ is not known at the time of credit scoring, so the different preferences of the lender and the regulator represent different trade-offs between the high and low states of the economy.
\end{application}

\begin{application}[Changing target populations in medical testing]
    \label{app:target}
    A hospital considers deploying an algorithmic screening tool to support testing decisions for its patients. The algorithm is optimized to maximize predictive power based on the vendor's baseline sample of patients, $U^A_\theta(f) = \E_\theta[u(f(x),y)|d{=}\text{vendor}]$.
    The hospital, however, cares about the performance of the screening tool on its specific group of patients, $U^P_\theta(f) = \E_\theta[u(f(x),y)|d{=}\text{hospital}]$. While the hospital's patients' health outcomes still follow the same conditional distribution $y|x$, they may have different demographics and medical histories $x$ than the baseline sample, and the hospital would therefore trade off net benefits across different patient groups differently than the vendor.
\end{application}

In all three examples, the preferences $U^P_\theta$ for the principal and $U^A_\theta$ for the agent are only partially aligned, since the principal has distributional preferences (over disparate impact, or ``fairness'' in our discussion to follow), cares about a different distribution of outcomes $y$ (due to risk preferences, or more generally what we refer to below as ``model shift''), or puts weight on different parts of the covariates $x$ (changes in the target population, or what we refer to below as ``covariate shift'').

Having set up and motivated misaligned preferences,
we now model the oversight over algorithms as a standard delegation problem where we allow for the principal to impose restrictions on the agent's choice of prediction functions.
Specifically, the principal first receives a signal $\pi$ in the form of a prior over the training signal $\theta \in \Theta$, chooses a restriction $\F_\pi \subseteq \F$ of functions $f$ the agent can choose from, and then the agent chooses a function $f \in \F_\pi$ upon learning the realization $\theta \sim \P_\pi$.
This setup follows the classical delegation-set approach without transfers of \cite{holmstrom1977} as further developed in \cite{melumad1991communication}, \cite{alonso_optimal_2008}, and \cite{Frankel2014-yn}.

To this standard model of delegation, we add a central concern in overseeing complex algorithms, namely that the functions $f \in \F$ may be too complex to be fully described to (or understood by) the principal.
Instead of formulating restrictions on $f$, we therefore assume that the principal can only restrict functions based on simpler descriptions (or ``explanations'') $\ex f \in \Ex$, so that the restrictions available to the principal take the form
\begin{align*}
    \F_\pi = \{f \in \F ; \ex f \in \Ex_\pi\}
\end{align*}
for some $\Ex_\pi \subseteq \Ex$. 
Crucially, the space $\Ex$ of model descriptions is typically smaller than the full space of functions $\F$, in which case the mapping $\ex: \F \rightarrow \Ex$ loses information because $f$ cannot be fully described.
For now, we take the set $\F$ of ex-ante permissible prediction functions as well as the explanation technology $\ex$ as given and later return to the implications of different choices of $\F$ and $\ex$.

Simple explanations of complex algorithms may in practice correspond to simple proxy models that project a prediction function onto a few key variables, variable importance scores, or a few interpretable key statistics.
For example, in our empirical example in \autoref{sec:empirical} we approximate a complex credit score based on boosted trees with over 500 covariates by a simple linear regression on a few key regressors.
Similarly, tools like SHAP \citep{lundberg2017unified} represent the main drivers of complex machine-learning predictions in terms of Shapley values associated with the contribution of individual covariates.
For random forests, variable importance measures similarly provide a simple description of each covariate's role in improving predictions.
In all of these cases, the simple descriptions represent an informative but incomplete summary of the output of complex algorithms.

We note that we use the term ``explanation'' and ``algorithm'' somewhat liberally here.
Specifically, what we call ``explanations'' are ex-post descriptions of the (potentially complex) prediction function $f$ chosen by the agent, which we interpret as the output of a machine-learning algorithm.
But we do not directly consider descriptions or explanations that extend from the function itself to how it was created by the underlying algorithm.
We discuss such extensions in \autoref{subsec:theory-extensions} below.

Embedding the complexity constraint into our delegation model,
we thus consider the following game:
\begin{enumerate}[nolistsep]
    \item The principal receives a signal $\pi$ in the form of a prior over the agent's training signal $\theta \in \Theta$, and chooses a restriction $\Ex_\pi$.
    \item The agent receives a training signal $\theta \sim \P_\pi$, and chooses a function $f_\theta \in \F$ subject to $\ex f \in \Ex_\pi$.
\end{enumerate}
In this model, the agent chooses $f_\theta \in \F_\pi = \{f \in \F; \ex f \in \Ex_\pi\}$ to maximize $U^A_\theta(f)$, while the principal chooses $\Ex_\pi$ to maximize the average expected utility $\E_\pi U^P_\theta(f_\theta)$.
If the explanation $\ex$ is exhaustive, such as in the case where $\Ex=\F$ and $\ex$ is the identity, we recover a standard delegation game.
Our main focus is instead on the case where $\ex: \F \rightarrow \Ex$ maps complex functions to simple explanations, leading to a loss of information because multiple prediction functions $f$ are described by the same simple explanation $\ex f$.

Before discussing concrete implementations of the simplicity constraint,
we note an alternative framing of the delegation model. Instead of being motivated by an information asymmetry between principal and agent, the model can describe a world where the agent's technology chooses functions so complex the principal cannot fully discern them, or the information constraint may reflect concerns about revealing proprietary technology.
In this case, the delegation problem remains applicable even when the principal has full information about $\theta$, and the principal chooses $\Ex_\theta$ to maximize $U^P_\theta(f_\theta)$.
By focusing on this case of known $\theta$, we aim to separate the implications of constraints on the complexity from those of private information.
Our general model remains applicable, however, to cases where both private information and asymmetries in the ability to comprehend fully complex prediction functions play a role.
Such cases may be particularly relevant when not only computational technology but also training data are controlled by the agent.

\subsection{Linear explanations with known state and quadratic loss}
\label{subsec:linearquadratic}

Having set up a general model of delegation with complexity constraints, we now solve the game for a concrete implementation.
Specifically, we consider the case of quadratic loss, simple linear explanations on a subset $S$ of covariates, and state $\theta$ known to the principal.
For this case, we illustrate concrete solutions in an example.
In \autoref{sec:empirical}, we then apply the same setup in a large-scale empirical exercise.

First, we assume that the agent strives to minimize mean-squared error $\E_\theta[(y - f(x))^2]$ (that is, has utility $u(f(x),y)= - (y-f(x))^2$).
This criterion represents a natural and frequently employed benchmark, which we see as a general-purpose proxy for prediction quality that allows us to derive explicit solutions.

\begin{assumption}
    \label{as:quadratic}
    The agent minimizes squared-error loss, $u(f(x),y) = - (y - f(x))^2$.
\end{assumption}

Second, we assume that simple explanations are given by linear regressions of the predictions $f(x)$ on a set of a few key variables $S$.
For now, we take that set as given, and discuss its optimal choice in \autoref{subsec:optimalexplainers} below.

\begin{assumption}
    \label{as:linear}
    The explanation mapping $\ex: \R^{\X} \supseteq \F \rightarrow \Ex = \R^S$ is a linear projection onto a set of $|S| < \infty$ covariates $x_S \in \R^S$ that are not collinear, $\ex f = \argmin_{\beta \in \R^S} \E_\theta[(f(x) - x_S' \beta)^2]$.
\end{assumption}

Third, we assume that the agent chooses from (potentially complex) functions that include at least linear regression.
Analogous to the explainers, we take this set $\F$ as given for now.

\begin{assumption}
    \label{as:functions}
    The function space $\F$ forms a real vector space that includes at least the linear functions $f(x) = x_S' \beta$ for $\beta \in \R^S$.
\end{assumption}

Finally, we focus on the case where there is no information asymmetry about the data distribution.
This assumption puts our attention instead on asymmetric information about complex prediction functions.

\begin{assumption}
    \label{as:known}
    The principal learns the state $\theta$, that is, $\pi$ puts point mass on a specific $\theta$.
\end{assumption}

These choices simplify the analysis substantially and allow us to make the role and specific structure of complexity constraints explicit.
Concretely, the principal now applies constraints based on a simple linear proxy model with covariates $x_S$, subject to which the agent makes his choice.
We can explicitly write the best choice of principal and agent in terms of linear projections.

\begin{proposition}
    [Constrained agent choice]
    \label{prop:constrained}
    Write $f^A_\theta = \argmin_{f \in \F} \E_\theta[(y - f(x))^2]$ for the agent's first-best choice from $\F$, $\beta^A_\theta = \ex f^A_\theta$ for its projection onto $x_S$, and $r^A_\theta(x) = f^A_\theta(x) - x_S' \beta^A_\theta$ for the remainder.
    Then an equilibrium of the game is given by the principal imposing the constraint $\ex f = \beta^P_\theta \in \R^S$ and the agent choosing $\hat{f}_\theta(x) = x_S' \beta^P_\theta + r^A_\theta(x)$, where $\beta^P_\theta$ maximizes $U^P_\theta(x \mapsto x_S' \beta^P + r^A_\theta(x))$ over $\beta^P$.
\end{proposition}

In other words, under the above assumptions, the agent follows the principal's instructions for the linear part of the prediction function corresponding to covariates $x_S$, and chooses the remaining part according to his first-best choice.

Before discussing the optimal choice of explainer variables, we apply the above result to a simple linear-regression example.
In this example, the role of fully complex prediction functions is played by fully interacted linear regression on two binary covariates, while the role of simple explanations is represented by their projection onto only one of the covariates.

\begin{figure}
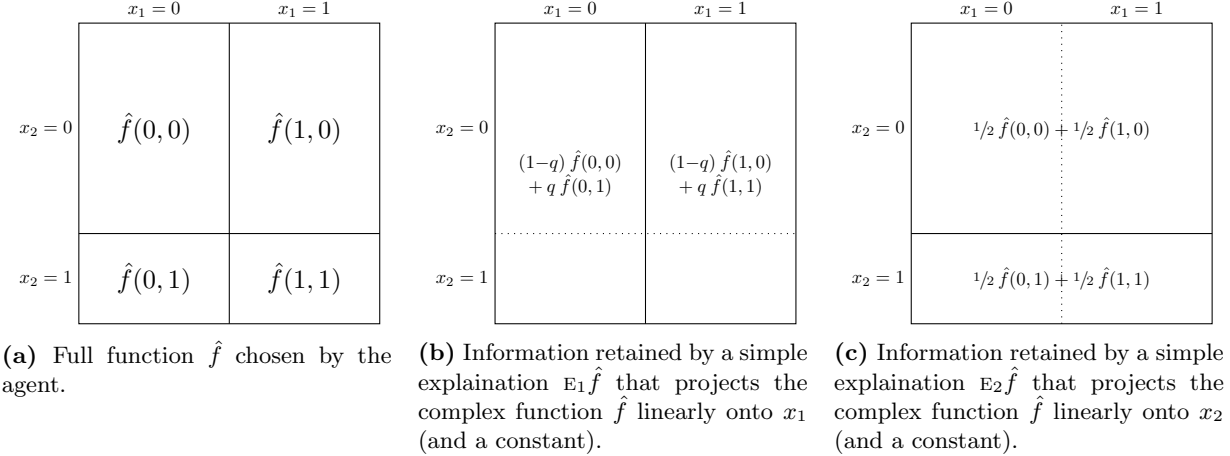

    \centering
    \begin{subfigure}[t]{0.3\textwidth}
        \centering
        \generalexplain
        \caption{Full function $\hat{f}$ chosen by the agent.}
        \label{fig:generalexplain}
    \end{subfigure}
    ~
    \begin{subfigure}[t]{0.3\textwidth}
        \centering
        \firstexplain
        \caption{Information retained by a simple explaination $\ex_1 \hat{f}$ that projects the complex function $\hat{f}$ linearly onto $x_1$ (and a constant).}
        \label{fig:firstexplain}
    \end{subfigure}
    ~
    \begin{subfigure}[t]{0.3\textwidth}
        \centering
        \secondexplain
        \caption{Information retained by a simple explaination $\ex_2 \hat{f}$ that projects the complex function $\hat{f}$ linearly onto $x_2$ (and a constant).}
        \label{fig:secondexplain}
    \end{subfigure}

    \caption{Illustration of the structure of a complex function $\hat{f}$ (left panel) as well as the information retained in simple explainers $\ex \hat{f}$ (center and right panels) from the \hyperref[ex:linearinteracted]{example}. Each cell corresponds to a combination of the values of the two binary covariates $x_1$ and $x_2$, and the values in the cells or across two cells represent the information retained in each case.}

    \label{fig:explain}
\end{figure}

\begin{example*}[name=Fully interacted linear regression,label=ex:linearinteracted]
    For concreteness, we consider the case of predicting an outcome $y \in \R$ (which can be binary) from a pair of binary covariates $x_1,x_2$, with the distribution being independent across $x_1$ and $x_2$ with $\E_\theta[x_1] = \sfrac{1}{2}, \E_\theta[x_2] = q \in (0, \sfrac{1}{2})$.
    In this world, we can without loss of generality write any (arbitrarily complex) conditional expectation of the outcomes as a fully interacted linear regression
    \[
        f_\theta(x) = \E_\theta[y|x] = \theta_0 + (x_1 - \sfrac{1}{2}) \: \theta_1 + (x_2 - q) \: \theta_2 + (x_1 - \sfrac{1}{2}) (x_2 - q) \: \theta_{12}
    \]
    since the covariates are both binary.
    However, we assume that the regulator can only understand projections of such functions onto one of the two covariates $x_1,x_2$ (and a constant $x_0$), $S = \{0,1\}$ or $S = \{0,2\}$ for the covariate space $\X = \{1\} \times \{1,0\}^2$.
    For $S = \{0,1\}$ as an example, for a function $\hat{f}(x) = \hat{\theta}_0 + (x_1 - \sfrac{1}{2}) \hat{\theta}_1 + (x_2 - q) \hat{\theta}_2 + (x_1 - \sfrac{1}{2}) (x_2 - q) \hat{\theta}_{12}$ the associated explainer $\ex_1: \R^{\X} \rightarrow \R^2$ yields $\ex_1 \hat{f} = \begin{psmallmatrix} \hat{\theta}_0 \\ \hat{\theta}_1\end{psmallmatrix}$.
    \autoref{fig:explain} illustrates both the information content of the full function $\hat{f}$ (left panel) as well as the information retained in the simple explainers $\ex_1 \hat{f}$ (center panel) for the case $q = .3$.
    Assuming that the agent's choices are otherwise unconstrained ($\F = \R^\X$), this means that the principal determines $\hat{\theta}_0, \hat{\theta}_1$ according to her preference (say, $\beta_\theta^P$), and the agent chooses $\hat{\theta}_2, \hat{\theta}_{12}$ according to his (that is, $\theta_2,\theta_{12}$), to obtain
    \[
        \hat{f}_\theta(x) = \underbracket{\beta_{\theta,0}^P + (x_1 - \sfrac{1}{2}) \beta^P_{\theta,1}}_{\textnormal{controlled by principal}} + \overbracket{(x_2 - q) \theta_2 + (x_1 - \sfrac{1}{2}) (x_2 - q) \theta_{12}}^{\textnormal{controlled by agent}} = x_S' \beta^P_\theta + r^A_\theta(x).
    \]
    As in the general case of \autoref{prop:constrained}, the choice of the residual $r^A_\theta(x) = (x_2 - q) \theta_2 + (x_1 - \sfrac{1}{2}) (x_2 - q) \theta_{12}$ does not depend on the principal's choice.
\end{example*}

In the remaining part of this section, we now apply the result from the example to the specific applications introduced in \autoref{subsec:modelsetup}.
In each application, we consider concrete choices of $x_1$ and $x_2$ in order to illustrate the structure of constrained solutions.

\begin{application}[name=Disparate impact in job screening,continues=app:DI]
    We assume for concreteness that job applicants are scored based on whether they have a college degree ($x_1$) and whether they have any gaps in their resume ($x_2$) in order to predict future on-the-job performance $y$.
    If the manager only reports a simple description of the screening score in terms of college degree, then in equilibrium the company has full control over how a college degree affects screening, but the manager's algorithm has full flexibility over the marginal influence of gaps in the resume on applicant scoring.
    If gaps in the resume are correlated with gender $g$, then the resulting scores may still have excess disparate impact.
\end{application}

\begin{application}[name=Varying risk preferences in credit provision,continues=app:risk]
    We assume that the probability of repayment is estimated from two binary variables, past default ($x_1$) and whether the applicant has a home-equity line of credit (HELOC, $x_2$).
    For our illustration, we let having a HELOC be a positive indication of creditworthiness in the high state of the economy, but a negative indication in the low state ($\theta_2(\text{high})> 0 > \theta_2(\text{low})$).
    We also allow the overall level of risk (intercept) to vary with the state of the economy, so that we can write
    \[
        \E_\theta[y|x,s] = \theta_0(s) + (x_1 - \sfrac{1}{2}) \: \theta_1 + (x_2 - q) \: \theta_2(s) + (x_1 - \sfrac{1}{2}) (x_2 - q) \: \theta_{12}.
    \]
    We assume that the future state $s$ of the economy at the time of potential repayment is unknown at the time credit scoring happens, but that the distribution over $\theta = (\theta(\text{high}),\theta(\text{low}))$ is learned in the training stage.
    In this case, the lender wants to evaluate HELOC in terms of its average effect, while the regulator prefers that HELOC is counted only as a negative.
    An explanation in terms of past default helps align the overall level of risk, but does not affect how the lender's algorithm leverages information about HELOCs.
\end{application}

\begin{application}[name=Changing target populations in medical testing,continues=app:target]
    The vendor's algorithm scores patients based on whether they have high blood pressure ($x_1$) and whether they have been diagnosed with a heart condition ($x_2$) to predict the incidence  $y$ of a disease.
    A simple explanation in terms of high blood pressure does not restrict how the vendor's algorithm scores the incremental information from a heart condition.
\end{application}

Across all applications, the simple explanation based on one of the two covariates gives the principal only partial control over the agent's algorithm.
Depending on the nature of misalignment and the choice of explainer covariates, the choices are only partially aligned.
While resulting prediction functions lead to better outcomes for the principal than not constraining the agent at all, they may still reflect substantial misalignment.

\subsection{Policy choice of simple models vs. simple explanations}
\label{subsec:policychoice}

Having laid out a simple delegation model as well as described its solution for quadratic loss and linear explainers,
we now consider different implications of the rules of the game itself, and how these restrictions affect the equilibrium outcome of the game.
We then provide formal results about when ex-ante restrictions to simple models can improve outcomes, and when oversight based on ex-post explanations dominates.

So far, we have taken the function class $\F$ and explainer mapping $\ex: \F \rightarrow \Ex$ as given.
In this section, we instead ask the ex-ante policy design question of how the function class $\F$ affects outcomes.
Under the assumptions of \autoref{subsec:linearquadratic}, we consider the difference in equilibrium outcomes between restricting $\F$ to be fully transparent, by which we mean an ex-ante restriction of $\F$ to simple linear functions on a subset of covariates, and leaving $\F$ unconstrained. 
In \autoref{subsec:optimalexplainers} below, we then ask how the design of the explainer mapping $\ex: \F \rightarrow \Ex$ affects equilibrium outcomes in the latter case, and solve for optimal explainers from the perspective of the principal.
Taken together, these two results allow us to characterize the optimal design of algorithmic regulation within our framework.

For this part of our analysis, we assume that principal and agent utilities take a similar form.
To this end, we introduce notation that is helpful to express important sources of misalignment between principal and agent.
We write $\mu_\theta$ for the probability measure over covariates $x$ implied by $\P_\theta$ and denote by $f_\theta(x) =  \E_\theta[y|x]$ the (possibly infeasible) prediction function that minimizes mean-squared error $\E_\theta[(y-f(x))^2]$ if there are no constraints on the functions $f$. Then 
\[U_\theta^A(f) = - \int_\X (f(x) - f_\theta(x))^2 \: \d \mu_\theta(x) + \const{}\]
with a constant part that does not depend on $f$, so the agent's utility is equivalent to minimizing $\int_\X (f(x) - f_\theta(x))^2 \: \d \mu_\theta(x)$.
This formulation allows us to express misaligned utility in terms of deviations in target functions and covariate distributions.

\begin{assumption}
    \label{as:dualrep}
    The principal's preference is equivalent to minimizing $\int_\X (f(x) - f^P_\theta(x))^2 \: \d \mu^P_\theta(x)$ for some target $f^P_\theta \in \R^{\X}$ and some probability measure $\mu^P_\theta$ over $\X$.
\end{assumption}

Hence, principal and agent utility may differ in the distribution over the target covariates and/or the target function.
A first version of misalignment is given by different target populations, holding the target function fixed across principal and agent.
Using terminology from machine learning, we call this variation across objectives ``covariate shift.''

\begin{definition}[Covariate shift]
    \label{def:covariateshift}
    Principal and agent utility differ by \emph{covariate shift} if $f^P_\theta = f_\theta$ but $\mu^P_\theta \neq \mu_\theta$.
\end{definition}

A second form of misalignment arises when populations are the same, but the agent cares about a different target function from that of the principal.
We call this change ``model shift'', as only the preferred prediction function changes.

\begin{definition}[Model shift]
    \label{def:modelshift}
    Principal and agent utility differ by \emph{model shift} if $\mu^P_\theta = \mu_\theta$ but $f^P_\theta \neq f_\theta$.
\end{definition}

In \autoref{prop:disparate} below, we also consider distributional preferences, for which the principal cares about average differences in predictions across groups as in \autoref{app:DI}.
There, we show that it can be interpreted as a special case of model shift.

We now contrast how two ex-ante policy options for the function class $\F$ and the explainer mapping $\ex$ can help align choices through setting the rules of the game laid out in \autoref{subsec:modelsetup}.
On the one extreme, we consider an ex-ante policy that restricts the function class $\F$ to functions that can be fully explained, so that $\F = \{x\mapsto x_{\Srestrict}'\beta; \beta \in \R^{\Srestrict}\} \cong \Ex$ for some set $\Srestrict$ of covariates. In that case, the principal has full control over the agent's choice, ensuring full alignment.
This policy option is one interpretation within our model of calls for fully transparent or explainable models.

On the other extreme, we consider an ex-ante policy that leaves the agent's choice fully unconstrained ($\F = \R^{\X}$) and only imposes constraints on the agent's choice via the explainer based on the covariates $x_{\Sexplain}$.
(Here, the explainer variables $x_{\Sexplain}$ do not necessarily have to be the same as those in the ex-ante constrained case $x_{\Srestrict}$.)
In this case, alignment is only partial.
This policy option connects to proposals that allow for fully complex functions, but subject these scores to targeted exams or require specific explanations for certification.

The following result describes the relative performance of full transparency and oversight based on ex-ante explainers for the specific cases of misalignment introduced above in terms of the principal's expected utility.

\begin{theorem}[Alignment through simplicity vs. alignment despite complexity]
    \label{thm:optimalregulation}
    If misalignment stems from \emph{covariate shift} with $\mu^P_\theta \ll \mu_\theta$, then choices from $\F = \R^{\X}$ are fully aligned, leading to higher principal and agent utility than any non-trivial ex-ante complexity constraint.
    If misalignment stems from \emph{model shift} then choices from $\F = \R^{\X}$ subject to an explanation constraint with explainers $\Sexplain$ lead to a higher expected principal utility outcome than ex-ante constraints with explainer covariates $\Srestrict$ if and only if
    \[
        \min_{\beta \in \R^{\Sexplain}} \int_\X (f_\theta(x) - f^P_\theta(x) - x_{\Sexplain}'\beta)^2 \: \d \mu_\theta(x) < \min_{\beta \in \R^{\Srestrict}} \int_\X (f^P_\theta(x) - x_{\Srestrict}'\beta)^2 \: \d \mu_\theta(x),
    \]
    that is, whenever the difference between targets is easier to explain than the principal's target itself.
\end{theorem}

In particular, considerations around different target populations only lead to misalignment for simple functions, but not when the agent is allowed to choose from a flexible function class.
(Here, we ignore  additional statistical considerations around resolving bias--variance trade-offs efficiently from limited data, in which case differential weighting of subpopulations may matter.)
With model shift, on the other hand, ex-ante simplicity restrictions can be beneficial to the principal when the difference between preferences is harder to explain than the first-best choice of the principal, but are also inefficient otherwise.

Covariate shift and model shift represent two specific sources of misalignment. We note that the latter of these comprises another type of misalignment between principal and agent where the principal cares about disparate impact, as in the hiring case of \autoref{app:DI} (and elaborated on in \hyperref[fn:DI]{Footnote~\ref{fn:DI}}).
Such a preference is equivalent to model shift with a specific target.
As a consequence, we can apply our previous result to the case of distributional preferences of the principal.

\begin{proposition}[Distributional preferences]
    \label{prop:disparate}
    Assume that the principal's utility differs by the agent's utility by a cost associated with lower average predictions of the minority group,
    \[U^P_\theta(f) = \E_\theta[u(f(x),y)] - \lambda (\E_\theta[f(x)|g{=}1] - \E_\theta[f(x)|g{=}0])\]
    for $g$ a binary indicator of group status (and $g=1$ the majority group).
    Writing $g_\theta(x) = \E_\theta[g|x]$ for the majority fraction at $x$ and $\bar{g}_\theta = \E_\theta[g]$ for the overall fraction of $g=1$, then the principal's utility is equivalent to model shift with $f_\theta^P(x) = f_\theta(x) - \frac{\lambda}{2 \bar{g}_\theta (1-\bar{g}_\theta)} (g_\theta(x) - \bar{g}_\theta)$.
    In particular, assuming that the explainer variables $x_{\Sexplain}$ include an intercept, choices from $\F = \R^{\X}$ subject to an explanation constraint with explainers $\Sexplain$ lead to a higher expected principal utility outcome than ex-ante constraints to only using covariates $\Srestrict$ if and only if    \[
        \frac{\lambda}{2 \bar{g}_\theta (1- \bar{g}_\theta)} \min_{\beta \in \R^{\Sexplain}} \int_\X (g_\theta(x) - x_{\Sexplain}'\beta)^2 \: \d \mu_\theta(x) < \min_{\beta \in \R^{\Srestrict}} \int_\X (f^P_\theta(x) - x_{\Srestrict}'\beta)^2 \: \d \mu_\theta(x),
    \]
    that is, whenever a $\frac{\lambda}{2 \bar{g}_\theta (1- \bar{g}_\theta)}$ fraction of the error in explaining group membership probabilities is smaller than the error in explaining the principal's target itself.
\end{proposition}

The result shows that distributional preferences are equivalent to model shift where the difference between principal and agent targets are given by an offset related to the relative fraction of majority and minority instances.
Equivalently, the principal prefers that predictions for the minority group are higher by a given offset than those of comparable majority instances.
The size of the offset depends on the relative size of groups as well as the strength of the disparate-impact preference.

This result is a direct generalization of a result in \cite{kleinberg_discrimination_2019} that considers optimal oversight over algorithms in a world where group identity $g$ is available to the agent. In this case, their result shows that optimal regulation enforces averages $\E[f(x)|g]$ across groups, while leaving predictions otherwise unconstrained. If $g$ is available as an explainer variable in $\Sexplain$, then this result directly follows from the proposition. Indeed, in this case, the expected loss on the left is zero, and ex-post explanations generically lead to higher expected utility than any ex-ante restriction.
Note, however, that our result is more general and does not require that the agent have access to group indicators, or that the final rule be allowed to depend on them.

Before discussing optimal choices of the covariate sets $\Srestrict$ and $\Sexplain$ (i.e., included regressors for the simple model and explainer variables for the complex model) we study the general result in the illustration from our example in \autoref{subsec:linearquadratic}, and apply it to the three cases laid out in \autoref{subsec:modelsetup}.
We start by comparing simple to complex solutions in the example.

\begin{example*}[name=Fully interacted linear regression,continues=ex:linearinteracted]
    We compare choices made by the agent from all functions $\R^{\X}$ subject to an explanation fixed by the principal to simple functions that are fully transparent.
    For $\Sexplain = \{0,1\}$ as above,
    in the former case the resulting function is
    \(
        \hat{f}_\theta(x) = \beta_{\theta,0}^P + (x_1 - \sfrac{1}{2}) \beta^P_{\theta,1} + (x_2 - q) \theta_2 + (x_1 - \sfrac{1}{2}) (x_2 - q) \theta_{12}
    \).
    If the agent is instead constrained to simple functions (here, $\Srestrict = \{0,1\}$, meaning a linear function regression on $x_1$ with intercept), the resulting function is
    \(
        \hat{f}_\theta(x) = \beta_{\theta,0}^P + (x_1 - \sfrac{1}{2}) \beta^P_{\theta,1}
    \),
    which the principal can now fully control.
\end{example*}

We next discuss the implications of the example across the three applications, which represent the three sources of misalignment we discuss above: distributional preferences in the job screening case, model shift for lending, and covariate shift in the medical application. 

\begin{application}[name=Disparate impact in job screening,continues=app:DI]
    We consider the choice of the company whether to restrict the hiring manager to score applicants only based on having a college degree ($x_1$). In that case, choices are aligned, but scoring may be inefficienct since it ignores the predictive information of gaps in the resume ($x_2$).
    If the misalignment over gaps in the resume is limited and the company's dislike of disparate impact is not too high, then a manager's algorithm that is allowed to be complex, but constrained in terms of its intercept and effect of college degrees, may still be preferred by the company.
\end{application}

\begin{application}[name=Varying risk preferences in credit provision,continues=app:risk]
    The regulator considers whether to allow the lender's algorithm to use HELOC ($x_2$), given the misalignment over that variable.
    If the regulator allows for complex algorithms that are only constrained in terms of overall risk level and how they use past default ($x_1$) via the explainer $\ex_1$,
    then choices over HELOC would be misaligned.
    From the regulator's perspective, this additional flexibility would only make sense if the benefit of including the interaction term outweighs the cost of the misaligned coefficient on HELOC.
\end{application}

\begin{application}[name=Changing target populations in medical testing,continues=app:target]
    If hospital and vendor only differ in the distribution of patients they care about, but not in their view on the incidence of the medical condition given available patient information, then unconstrained choices are not misaligned, no matter whether an explanation constraint is enforced or not.
    If, however, the vendor is barred from leveraging diagnosed heart conditions ($x_2$) in his algorithm, then choices are not only inefficient, but also potentially misaligned.
    This is because different joint distributions of high blood pressure ($x_1$) and heart condition across target populations imply different adjustments for the included coefficients $\theta_0,\theta_1$ to account for the excluded coefficients $\theta_2, \theta_{12}$ (equivalent to the dependence of omitted variable bias on the covariate distribution).
\end{application}

For the choice between ex-ante restrictions and flexible prediction functions with ex-post explanations, this result makes a stark prediction: unless misalignment in targets is of at least the same order of magnitude and complexity as the principal's target function itself, complexity restrictions are not optimal for the principal, and will hurt both agent and principal utility when we compare restrictions and explanations using the same covariates.
In the specific case of a preference for parity across groups, complexity restrictions are suboptimal unless the preferences $\lambda$ against disparity is severe \emph{and} group membership is hard to explain. 
In principle, a combination of both policy levers can further improve outcomes, but may be unrealistic in practice.

\subsection{Optimal model explanations}
\label{subsec:optimalexplainers}

In the previous sections, we provided a characterization of when oversight of flexible functions based on ex-post explanations can improve over ex-ante restrictions that limit the agent to explainable models.
We now discuss the optimal choice of explainers that ensure second-best outcomes from the perspective of the principal.
This addition can be interpreted as extending our model in \autoref{subsec:modelsetup} by an initial stage at which the principal decides on the form of the required explanation.
Practically, this means that the principal also specifies the set of covariates to include in the simple description of the complex model.

A standard approach to explaining prediction choices by the agent would be to leverage an explanation mapping that can capture the maximal amount of information about the agent's prediction function, which we call the \emph{misalignment-agnostic explainer}.
In our specific setting, this would be the mapping $\ex_0 f = \argmin_\beta \int_{\X} (f(x) - x_{S_0}'\beta)^2 \d \mu_\theta(x)$ with $S_0 = \min_{S \in \mathcal{S}} \min_\beta \int_{\X} (f_\theta(x) - x_{S}'\beta)^2 \d \mu_\theta(x)$ chosen to minimize the discrepancy between explained and unexplained part of the agent's first-best prediction function $f_\theta$, subject to the complexity constraint $|S| \leq k$.

However, this agnostic explainer is not generally optimal.
Instead of relying on a set of covariates that best explain the average behavior of the prediction function, the following result shows that the principal can do better by requiring an explanation optimized for the specific source of preference misalignment.

\begin{theorem}[Targeted explainer]
    \label{thm:optimalexplainer}
    If misalignment stems from \emph{model shift} and the function class is unconstrained, $\F = \R^{\X}$, then
    the optimal explainer solves
    \[
        S^*
        =\min_{S \in \mathcal{S}} \min_\beta \int_{\X} (f_\theta(x) - f_\theta^P(x) - x_{S}'\beta)^2 \d \mu_\theta(x),
    \]
    that is, the optimal explainer is \emph{targeted} to the misalignment between principal and agent.
\end{theorem}

Here, we assume that the principal knows the true distribution $\P_\theta$ of the data when deciding on explainer variables. The results in this section generalize to the case where the principal only has some belief about $\theta$ when specifying required explainer variables, such as the prior $\pi$ from \autoref{subsec:modelsetup} or a belief coming from a less informative hyper-prior. In this case, the minimization is over the expectation $\E \min_\beta \int_{\X} (f_\theta(x) - f_\theta^P(x) - x_{S}'\beta)^2 \d \mu_\theta(x)$ with respect to that belief.

The targeted explainer takes a particularly simple form when the source of model shift stems from different distributional preferences as in \autoref{prop:disparate}.
Specifically, the optimal explainer can be obtained from a simple prediction problem.

\begin{corollary}[Targeted explainer for disparate impact]
    \label{cor:optimaldiexplainer}
    If misalignment stems from \emph{distributional preferences} and the function class is unconstrained, $\F = \R^{\X}$, then
    the optimal explainer solves
    \[
        S^*
        =\min_{S \in \mathcal{S}} \min_\beta \E_\theta[ (g - x_{S}'\beta)^2],
    \]
    that is, the optimal explainer is a best explainer of the majority indicator.
\end{corollary}

In other words, the covariates that best align choices with the principal's distributional preferences are those indicative of group membership, rather than those most predictive of the outcome itself.
If group membership itself is available to the agent, then we recover the result from \cite{kleinberg_discrimination_2019}: in this case, the principal should regulate the algorithm based on group-specific averages $\E[f(x)|g]$, corresponding to an explainer that describes predictions by a regression on $g$, and leave the algorithm otherwise unconstrained.
However, our result also extends to cases where the agent does not have access to protected group characteristics or is barred from using them, in which case the optimal explainer amounts to describing the algorithm in terms of features of the data that are related to differences across groups.

Equipped with the notion of a targeted explainer, we can now compare the ex-ante policy choice from \autoref{subsec:policychoice} between restricting models ex-ante to some set $\Srestrict$ or leaving models unrestricted and choosing explanation covariates $\Sexplain$, where we assume that the respective sets are chosen optimally from the perspective of the principal.

\begin{corollary}[Optimal regulation]
    \label{cor:optimalregulation}
    If misalignment stems from \emph{covariate shift} with $\mu^P_\theta \ll \mu_\theta$, then leaving the function class ex-ante unconstrained is optimal from both agent and regulator perspectives, and explainer choices are inconsequential.
    If misalignment stems from \emph{model shift} and restrictions are chosen optimally, then leaving the function class ex-ante unconstrained is preferred by the principal if and only if
    \[
        \min_{\Sexplain \in \mathcal{S}} \min_\beta \int_\X (f_\theta(x) - f^P_\theta(x) - x_{\Sexplain}'\beta)^2 \: \d \mu_\theta(x) < \min_{\Srestrict \in \mathcal{S}} \min_\beta \int_\X (f^P_\theta(x) - x_{\Srestrict}'\beta)^2 \: \d \mu_\theta(x),
    \]
    where the targeted explainer from \autoref{cor:optimaldiexplainer} is optimal in the unconstrained case.
\end{corollary}
    
We close our discussion of the model results by applying them back to the three motivating applications within our simple linear example, in which we contrast the agnostic and targeted explainers.

\begin{example*}[name=Fully interacted linear regression,continues=ex:linearinteracted]
    We continue our simple example with two binary covariates $x_1,x_2$, and we assume that the cost of excluding each of the covariates is ex-ante equal.
    In this case, the misalignment-agnostic explainer focuses on the projection $\ex_1$ onto $x_1$ (and an intercept), since $x_1$ varies more than $x_2$ and thus provides more information about the agent's choice.
    However, if choices are aligned along $x_1$, but not along $x_2$, then the targeted explainer instead regresses the agent's choice onto $x_2$ (and an intercept).
    The resulting explainer $\ex_2$ is represented in the right panel of \autoref{fig:explain}.
\end{example*}

Within our applications, we illustrate how targeted and agnostic explainers may differ in practice.
In \autoref{sec:empirical}, we provide an analogous illustration from an empirical application to consumer lending.

\begin{application}[name=Disparate impact in job screening,continues=app:DI]
    Assuming that choices are misaligned over the use of gaps in the resume ($x_2$) only, then regulating based on the explainer $\ex_2$ that is targeted toward the misalignment aligns choices.
    The targeted explainer $\ex_2$ also has an intuitive interpretation in this case: it describes the role of those covariates that are most related to differences across groups.
    On the other hand, the misalignment-agnostic explainer $\ex_1$ (based on college degree, $x_1$) is better at capturing the overall behavior of the scoring function, but worse at capturing the aspect that is relevant for disparate impact with respect to gender.
\end{application}

\begin{application}[name=Varying risk preferences in credit provision,continues=app:risk]
    The targeted explainer $\ex_2$ fully captures the source of misalignment, which here is the overall level of risk as well as the role of HELOCs in credit scoring ($x_2$).
    The impact of past default (and the interaction of both) is left unrestricted, which happens to be optimal in this case.
    The targeted explainer again permits an intuitive interpretation: it captures those features that are most related to differences across economic states $s$.
\end{application}

\begin{application}[name=Changing target populations in medical testing,continues=app:target]
    Since choices are aligned without any restrictions in the case of different target populations,
    the choice of explainer covariates does not affect the outcome.
    Hence, there is only a cost to simplicity in this example.
    However, if the hospital were to force the vendor to exclude one of the covariates, then keeping the variable that is most important in the patient distribution of the hospital would be the better among bad options.
\end{application}

While these illustrations are extreme in that they allow for perfect alignment, we explore empirically below how the trade-off between simple and complex functions as well as agnostic and targeted explainers play out in data from the context of consumer lending, both for disparate-impact concerns as in \autoref{app:DI} and for divergent risk preferences as in \autoref{app:risk}.

\subsection{Variants and extensions}
\label{subsec:theory-extensions}

In our model, we have so far illustrated trade-offs in algorithmic regulation in a stylized setup that focuses on tractability.
Here, we discuss extensions.

\paragraph{Double-selection explainers for disparate impact.}

In \autoref{app:DI} and \autoref{prop:disparate}, we present results on aligning choices when the source of misalignment is the principal's dislike for disparate impact.
The targeted explainer takes an intuitive form: its features represent the set of covariates that best describes group membership.
This specific result is driven by the linear form of the penalty in the principal's utility function (see also \hyperref[fn:DI]{Footnote~\ref{fn:DI}} for a discussion).
Different forms of the penalty, such as a squared penalty, may affect the optimal design of the explainer. 
Specifically, it could be optimal to choose those features that are important for \textit{both} group membership \textit{and} the outcome model.
This has a connection to the literature on double selection in machine learning \citep[and specifically][]{belloni2014inference}, which emphasizes the selection of all potential confounders that are related to the assignment of treatment \textit{or} the outcome in order to avoid selection mistakes.

\paragraph{Audits based on realized outcomes.}

We assume in our model that restrictions are set on prediction functions before they are deployed and before their consequences are realized, such as employment, credit, and testing decisions. Once the actual decisions and their consequences are measured, audits may effectively uncover \emph{realized} misaligned choices.
For example, managers may lose their bonuses for unequal hiring decisions, banks may be fined for experiencing excess defaults, and vendors may be punished for misallocated tests.
While such information is often only available with a delay, it could improve adherence at the training stage.
Optimal regulation would then combine restrictions on model explanations that can be verified before deployment, and fines for bad outcomes.
However, as long as outcomes are uncertain and agents risk averse, an optimal mix of regulatory instruments would include the use of explainers, since regulating only based on ex-post outcomes in those cases
cannot distinguish well between bad choices and bad circumstances.
As a result, consequence-based regulation may be slow, have limited effect, or lead to overly conservative agent choices.
At the same time, the availability of realized consequences may further limit the need for ex-ante complexity constraints.

\paragraph{Direct regulation of binary decisions.}
In our model above, we assume that preferences are over real-valued predictions, such as a hiring manager's assessment, a lender's credit score, or a health provider's risk assessment.
We may instead apply our model directly to the decisions taken by the agent, and focus on binary policies that determine who is hired, who receives credit, or who is tested.
In those cases, we would expect similar results as in the more tractable continuous case, with the inefficiency of ex-ante complexity restrictions depending on cases in a neighborhood of the agent's decision threshold.

\paragraph{Ex-post vs. process-based descriptions.}
Our discussion of algorithmic regulation has focused on describing resulting prediction functions, rather than regulating the procedure by which training data and machine-learning pipelines yield prediction functions.
Considering the entire process beyond ex-post descriptions of resulting decision functions \citep[as, e.g., considered by][]{kleinberg2018discrimination} may be a necessity when it is relevant for legal and regulatory considerations and could provide additional tools for making algorithmic decisions more transparent without having to rely on inefficient restrictions to complexity.

\section{Empirical Implementation}
\label{sec:empirical}

We build an empirical counterpart to our model using machine-learning predictors subject to constraints. In this large-scale empirical illustration, 
we evaluate the different regulatory approaches from our model in the context of algorithmic credit scoring.
We focus on two cases of preference misalignment between a financial regulator (principal) and a lender (agent).
In the first case, the regulator has a taste for more equal outcomes across racial or ethnic groups -- or a taste for less disparate impact.
In the second case, the regulator cares about model fit in an economic downturn with elevated default rates, similar to a stress-test scenario. 
Across both cases, we demonstrate the benefit of allowing for complex prediction functions and the value of using targeted explainers.

\subsection{Data setup}

Our base dataset is a random sample of about 300,000 credit bureau files that have a newly opened credit card in 2012. 
We focus on credit cards since they are a widely used credit product for which algorithmic underwriting is already in use by some providers. 
This sample is drawn from the larger credit bureau panel constructed in \cite{blattner_laura_how_2021}, which used a probability-weighted sampling strategy to ensure sufficient representation from minority groups; this feature is important in our disparate impact empirical application. 
We draw another random sample of 50,000 credit bureau files from the same credit bureau panel for consumers who were rejected for a recent credit card application; these data are useful for our empirical application focused on misaligned preferences over borrower risk.
We split the data into a training dataset (80\% of the sample) and a test dataset (20\% of the sample).

Our main prediction outcome is credit card default of any severity up to 24 months after the card was opened. We capture this outcome by a repayment indicator $y \in \{1,0\}$, where $y=1$ denotes repayment and $y=0$ denotes default. The dataset contains several hundred credit report attributes, including detailed default histories, debt balances, utilization, credit report inquiries, and current and past debt obligations. For each variable, we create outlier and missing value flags and include these in the set of variables to predict default. We transform all balance variables into logs. We obtain a total of 518 variables. \autoref{tbl:descriptives} shows summary statistics for repayment as well as the ten variables with the highest importance value across a set of baseline prediction models of credit card default trained using XGBoost, random forest, elastic net, and neural net algorithms (see \autoref{tbl:performance} for performance summaries).
For our prediction exercises, we normalize all of the credit bureau variables so that in the main dataset of 300,000 opened credit cards, the variables are centered at zero with standard deviation equal to one.

In addition to the credit report attributes, we use an indicator for whether an individual belongs to a racial or ethnic minority. We draw on prior work in \cite{blattner_laura_how_2021} which computes minority/non-minority status following the industry-standard BISG methodology using name and geographic information to predict race and ethnicity (see \citealp{blattner_laura_how_2021}, for extensive validation of this imputation measure using a merge with HMDA data).\footnote{BISG uses a two-step procedure to assign race and ethnicity. First, the procedure follows an approach developed by \cite{sood_predicting_2018}. Their approach is implemented in a Python package available at \href{https://github.com/appeler/ethnicolr}{github.com/appeler/ethnicolr}. They model the relationship between the sequence of characters in a name and race and ethnicity using Florida Voter Registration data. After implementing this approach, the procedure updates each individual's baseline racial/ethnic probabilities with the racial and ethnic characteristics of the census block associated with her or his place of residence in 2000 using Bayes' Rule, and then updates the Bayesian posterior again using an individual's 2010 address and the 2010 census data. An individual is assigned to a racial/ethnic category if this category has the highest posterior probability for that individual. This two-step method is similar to methods used by the CFPB to construct race and ethnicity in fair lending analysis. Work by the \cite{cfpb_using_2014} shows that combining geographic and name-based information outperforms methods using either of these sources of information alone. BISG classification errors can be correlated with economic disadvantage, as minorities who live in predominately non-minority geographies are more likely to be misclassified by BISG as non-minority \citep{blattner_laura_how_2021} and are more likely to have higher education and income \citep{greenwald2023regulatory}. Such issues are important in some contexts but are unlikely to qualitatively affect our conclusions in this setting.}
We aggregate all ethnic and racial minorities into a single minority category and define its complement as the non-minority (or majority) group. Given the probabilistic sampling strategy in \cite{blattner_laura_how_2021}, the share of minority applicants in our sample is about 19.7\%.

\subsection{Lender and regulator preferences}

Across both applications, the role of the principal is played by a financial \textit{regulator} who cares either about financial stability or fair and non-discriminatory lending practices.
The agent in this implementation of our game is a \emph{lender} who uses past data on credit histories and defaults to score future loan applicants.

To be consistent with our theoretical approach, we provide results for the case where loss functions are mean-squared error in predicting repayment, $u(f(x),y) = - (y - f(x))^2$ as in \autoref{as:quadratic}.
We also report implications for outcomes when these scores are used to approve a given fraction of applicants.
We think of the mean-squared error approach as proxying for regulator and lender utility in a way that captures the quality of credit scoring beyond approval decisions. For example, scores may matter not only for approvals, but also for credit terms or for approvals across a range of thresholds.

In the disparate-impact example, the lender's preferred model is simply the best fit of credit repayment given background characteristics of the applicant. The regulator additionally penalizes differences in average repayment predictions across majority and minority applicants.
Hence, as in \autoref{app:DI},
\begin{align}
    U^A_\theta(f) &= - \E_\theta[(y-f(x))^2]
    \tag{DI-A}
    \label{eq:DI-A}
    \\
    U^P_\theta(f) &= - \E_\theta[(y-f(x))^2]
    - \lambda^* \: (\E_\theta[f(x)|g{=}\text{majority}] - \E_\theta[f(x)|g{=}\text{minority}])
    \tag{DI-P}
    \label{eq:DI-P}
\end{align}
We calibrate the penalty $\lambda^*$ such that for the most flexible model, the correlation of the output with the minority flag is zero.
Throughout, we assume that group identity $g$ does not enter the credit scoring function $f$ directly.

For the risk case we assume that the regulator's and the lender's preferences place different weights on good (or ``high'') and bad (or ``low'') states of the economy, with the regulator particularly concerned about the low state, similar to a stress-test scenario. We suppose that a particular group of borrowers -- the sample of 50,000 individuals recently rejected for a credit card application -- are sensitive to the state of the economy and default with probability 1 in the low state and probability 0 in the high state. We then refer to the lender's credit score as having a low-state MSE and a high-state MSE. For purposes of illustration, we assume that the regulator puts weight only on the low-state MSE, while the lender puts $\ell^* = 20\%$ weight on the low-state MSE. 
As in the illustration in \autoref{app:risk} above, the
lender (agent) and regulator (principal) utilities are therefore given by
\begin{align}
    U^A_\theta(f) &= - \E_\theta[(y-f(x))^2] = - (1-\ell^*) \: \E_\theta[(y-f(x))^2|s{=}\text{high}] - \ell^* \: \E_\theta[(y-f(x))^2|s{=}\text{low}],
    \tag{R-A}
    \label{eq:R-A}
    \\
    U^P_\theta(f) &= - \E_\theta[(y-f(x))^2|s{=}\text{low}].
    \tag{R-P}
    \label{eq:R-P}
\end{align}
Here, we assume that the (future) state is not known when (current) credit scoring decisions are taken, so that the state $s$ does not enter $f$.

In addition, for expositional purposes, we also consider intermediate preferences that interpolate between the regulator's and lender's preferences and show a Pareto frontier of achievable utility.
In the disparate impact case, we vary the weight $\lambda$ put on disparate impact, varying the parameter from zero (lender utility) to the calibrated value $\lambda^*$ (regulator utility),
showing the tradeoff between mean-squared error and disparate impact.
In the risk application, we vary the weight $\ell$ put on the low state from $0\%$ to $100\%$, where $\ell^*=20\%$ represents lender utility, and $100\%$ represents the preference of the regulator.

\subsection{Optimal explainers and constrained solutions}

To implement explainability restrictions, we assume that the regular chooses a set $S^*$ of $|S^*|=k$ covariates to use in an explainer. We set $k=5$ in our main implementation, and we also present robustness results for $k=10$ and $k=20$. 
The regulator then constrains the lender to choose only among functions $f: \X \rightarrow \R$ for which the regression coefficients $\ex f$ in a linear regression of $f(x)$ on the variables in $S^*$ (and an intercept) agree with a target value $\beta^*$ chosen by the regulator.
By \autoref{prop:constrained}, the optimal coefficient $\beta^*$ is the linear regression of the regulator's first-best solution on $S^*$ (and an intercept).
Likewise, by \autoref{thm:optimalexplainer}, the set of $k$ optimal explainer covariates is chosen so that it minimizes mean-square error in a linear regression relating all data features to the difference between the regulator's and the lender's preferred prediction function.

In our two applications, the optimal covariate sets for targeted explainers take a particularly simple form.
For the disparate-impact case, the optimal set of covariates solves 
\begin{align}
    \label{eq:DI-Ex}
    \tag{DI-Ex}
    \argmin_{S \in \mathcal{S}} \min_\beta \E_\theta[ (g - x_S'\beta)^2]
\end{align}
where $g \in \{1,0\}$ denotes group identity (with $1$ corresponding to the majority group).
Meanwhile, for the risk application, writing $h \in \{1,0\}$ for those applicants whose default is sensitive to the state of the economy,
the optimal set solves
\begin{align}
    \label{eq:R-Ex}
    \tag{R-Ex}
    \argmin_{S \in \mathcal{S}} \min_\beta \E_\theta[ (h - x_S'\beta)^2].
\end{align}

We contrast these two targeted explainers with a misalignment-agnostic explainer that simply explains as much as possible of the overall variation in repayment,
namely
\begin{align}
    \label{eq:Ex}
    \tag{Ex}
    \argmin_{S \in \mathcal{S}} \min_\beta \E_\theta[ (y - x_S'\beta)^2].
\end{align}
Except for differences in the estimation sample we choose for our two applications, the misalignment-agnostic explainer is the same across the two settings.
Throughout, we include (but do not penalize) intercepts.

We implement these solutions empirically using boosted trees, where we separate training from evaluation in order to ensure valid inference.
Details on optimization and sample splitting are presented in \apxref{apx:optimization}.

\subsection{Empirical results}

\begin{figure}[h]
     \centering
     \begin{subfigure}[b]{0.45\textwidth}
         \centering
         \includegraphics[width=\textwidth]{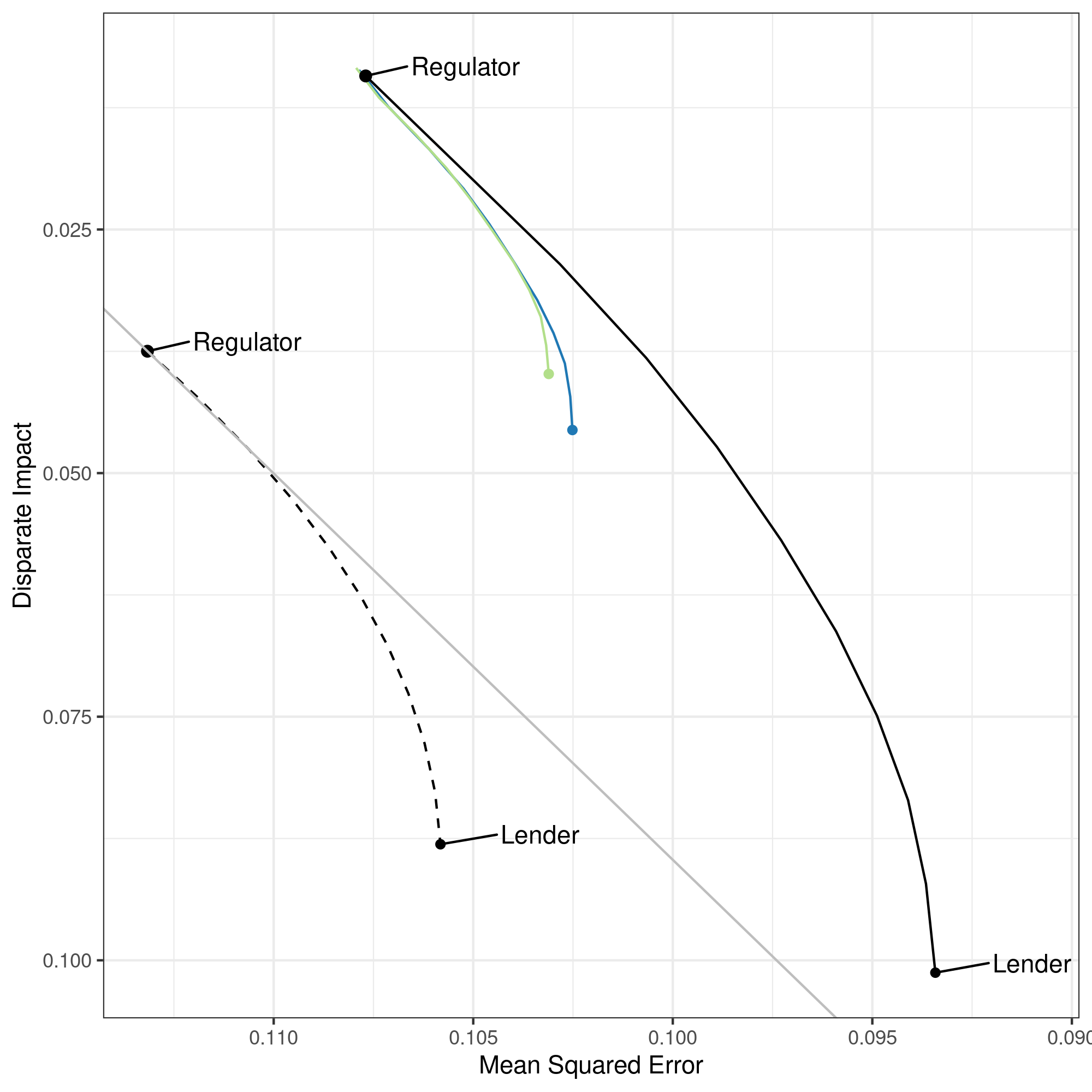}
         \caption{DI application}
         \label{fig:results_di}
     \end{subfigure}
     \hfill
     \begin{subfigure}[b]{0.45\textwidth}
         \centering
         \includegraphics[width=\textwidth]{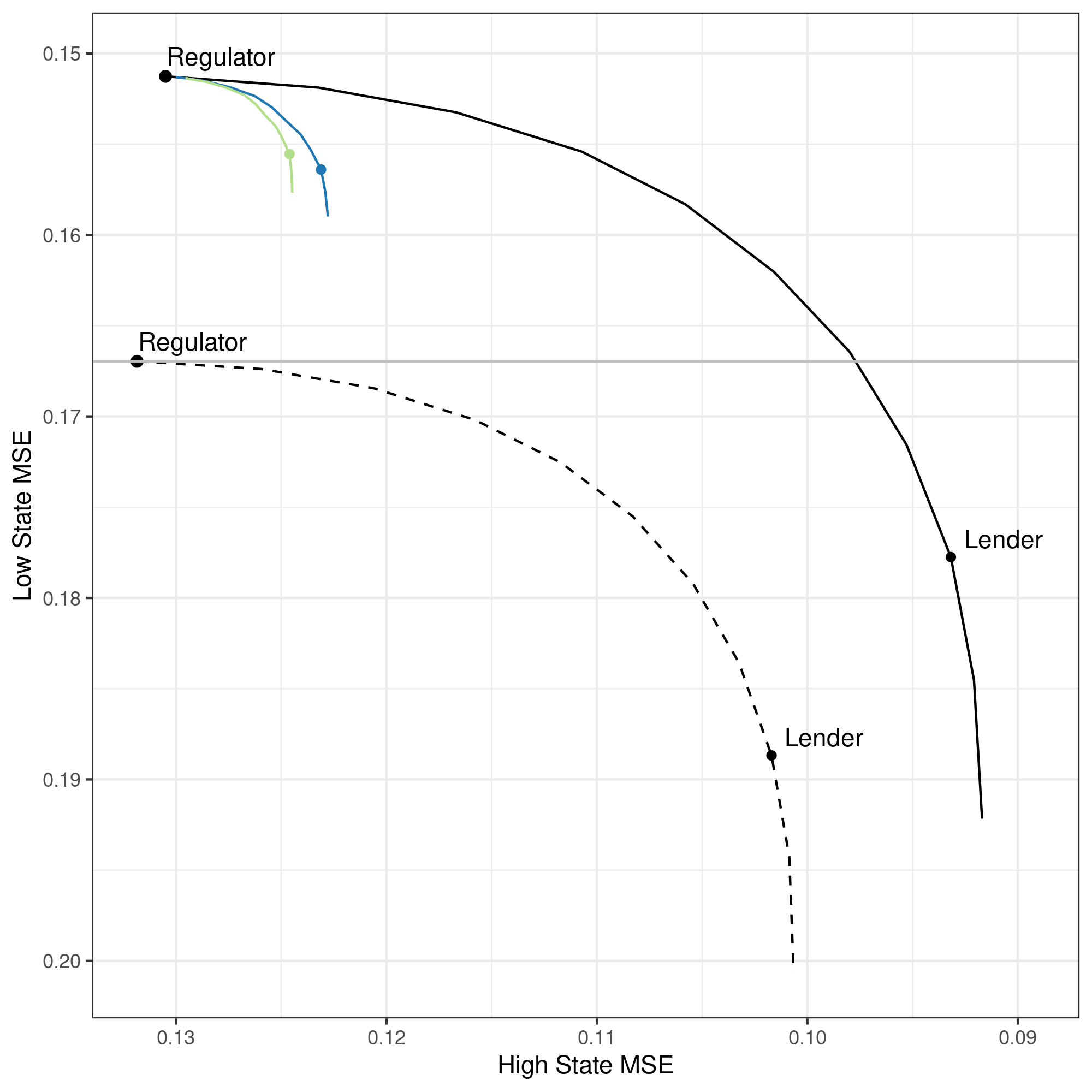}
         \caption{Risk application}
         \label{fig:results_risk}
     \end{subfigure}

     \begin{subfigure}[b]{\textwidth}
         \centering
         \includegraphics[width=\textwidth]{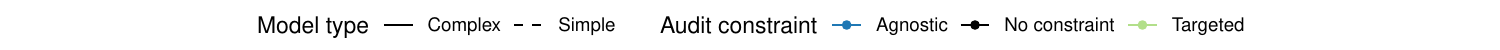}
     \end{subfigure}     
        \caption{
        Out-of-sample performance of unconstrained and constrained prediction models across key objectives for the disparate impact (left) and risk (right) applications.
        Complex models are XGBoost models on all 518 covariates, while simple models are linear regression on five covariates chosen by the LASSO.
        The frontiers vary the relative weight put on each objective, where the regulator and lender marks correspond to the solutions maximizing the empirical analog to the objectives \eqref{eq:DI-P}, \eqref{eq:DI-A} and \eqref{eq:R-P}, \eqref{eq:R-A} respectively.
        The colored lines represent prediction models subject to constraints imposed by the agnostic and targeted explainers, respectively.
        The gray line represents the points at which the regulator would be indifferent to its preferred simple model.}
        \label{fig:results}
\end{figure}

The main empirical results for both applications are summarized in \autoref{tbl:results} and \autoref{fig:results}.
\autoref{fig:results} illustrates the different components of utility that the regulator and lender optimize for. \autoref{tbl:results} evaluates the various credit scores by examining implied credit approval decisions, assuming that all applicants with a credit-score-predicted repayment probability of at least 85\% are approved for a loan.
Throughout, results for the disparate-impact application are listed in panel (a), and results for the risk application in panel (b).

We first consider results for the disparate impact application, for which we compare simple, fully explainable models to more complex ones. 
\autoref{fig:results_di} represents the trade-off between the fit of the model on the $x$-axis (which the lender wants to maximize) and disparate impact on the $y$-axis (which the regulator trades off with fit).
The different preferences are exemplified by the first-best lender and regulator solutions; the lender's first-best solution is towards the right, while the regulator's first-best solution achieves lower disparate impact at the cost of some fit.
Ignoring explainer constraints for the moment, we note that the more complex XGBoost models (solid black line) Pareto-dominate the less complex, fully explainable linear models (dashed lines), in the sense that there are more complex models that have better fit and lower disparate impact than either the simple regulator or lender models.
Moving towards more complex models thus has the potential to achieve better outcomes for both regulator and lender.
At the same time, the \textit{unconstrained} model the lender would choose among all complex models (bottom right lender solution) has high disparate impact, and would only represent a marginal improvement over the regulator's preferred simple, fully explainable model.
The main question is, therefore, whether the regulator can impose constraints based on simple explanations so that allowing the lender to choose from complex credit scores leads to a clear improvement in the regulator's utility.

To study this question within the disparate-impact application, we consider policies that allow the lender to choose among complex models subject to a constraint on simple explanations based on five explainer variables only.
The blue line in \autoref{fig:results_di} represents the Pareto frontier of complex models subject to a constraint based on the agnostic explainer.
The agnostic explainer captures the maximal amount of information about the overall variation of the underlying credit-scoring model, and imposing a constraint based on it leads to higher regulator utility than allowing the lender to make unconstrained choices.
At the same time, the regulator can further improve disparate impact by instead constraining explanations based on a targeted explainer. The green line represents the Pareto frontier of complex models subject to a constraint based on the targeted explainer, and this frontier dominates that from the agnostic explainer.
This targeted explainer focuses on aspects of the credit score that are particularly informative about disparate impact and, thus, about the source of misalignment in our example.

In our analysis so far, we have focused on fit and disparate impact of the credit scoring function itself. In \autoref{tbl:results_di} we also consider disparate impact of binary loan approval decisions based on the credit score.
The ``DI on acceptance'' column reports the disparate impact of credit approval decisions when all applicants with a predicted repayment probability of 85\% or more are approved for credit.
These results confirm that imposing constraints on explanations can reduce disparate impact with only a small cost in predictive power.
While the unconstrained lender solution leads to a 4.5 percentage point higher approval rate among majority applicants, that gap is reduced to 0.9 percentage points with an agnostic explainer, and approximately zero when the specific targeted explainer is used.
The complex solution subject to a targeted explainer constraint thus outperforms the regulator's preferred simple, fully explainable model even in terms of disparate impact.
Meanwhile, the DI and MSE columns report the same results as those in \autoref{fig:results_di} for the lender's and regulator's unconstrained and simple models, as well for the lender's constrained solutions.

We next consider results for the risk application, finding overall similar patterns.
\autoref{fig:results_risk} shows the superior performance of the complex model and the value of explanation constraints: we ask whether all possible convex combinations of the regulator's and lender's preferences would prefer the complex model over a model that is constrained ex-ante to be simple. We trace out a frontier in the space of low-state utility and high-state utility that shows the performance of the complex model for all combinations of regulator and firm preferences. We then trace out analogous curves when the complex model is subjected to an agnostic explainer and a targeted explainer, and finally, analogous curves for the ex-ante simple model (with no explainer). As in the disparate impact case, we see that complex models constrained by either explainer are preferred by both the regulator and the firm to any simple model chosen by any weighted combination of regulator and firm preferences. Constraints play a central role in this result, as the unconstrained complex model is worse for the regulator than the regulator's preferred simple model. 

\autoref{tbl:results_risk} provides further details for the risk application. The first four rows show statistics for the lender's complex model when the regulator uses no constraint, an agnostic explainer, an optimal targeted explainer, and an ex-ante constraint to use simple models only; the next two rows show the regulator's preferred complex model and the regulator's preferred simple model. Across columns, the table shows model behavior in terms of false positives (accepted defaults), true positives (accepted non-defaults) and the mean-squared error across both defaults and non-defaults. These statistics are presented for both the high state and the low state; recall that the lender places less preference weight on low-state outcomes than the regulator does.

The table confirms that the different models perform as expected: the constrained models perform better in the low state than the unconstrained models, though not as well as the regulator's preferred model; models that perform better in the low state face a trade-off in that they perform worse in the high state; and complex models generally outperform simple models.

The results across both applications present an empirical illustration of our main theoretical results.
First, as in \autoref{thm:optimalregulation}, a complex model with an explainer outperforms a model that is constrained ex-ante to be simple in our data. Here, this holds both for the regulator's preferences and the lender's preferences; both sides are better off using the more complex model with an appropriate explainer.
Second, the choice of explainer matters, with the targeted explainer from \autoref{thm:optimalexplainer} providing better results for the regulator across both applications.
The empirical application therefore expands our illustration from the case of two binary variables in the \hyperref[ex:linearinteracted]{example} from \autoref{sec:model} to a realistic exercise in credit underwriting on over 500 covariates.
In \autoref{fig:complexity} in the appendix, we show that these results are robust to a larger number of explainer covariates than the five chosen for illustration in this section.

\begin{table}[h]
    \centering
    \begin{subtable}{\textwidth}
        \centering
        {\scriptsize	
        \begin{tabular}{l
    l
    S[table-format=1.4]
    S[table-format=1.4]
    S[table-format=1.4]
    S[table-format=1.4]
    S[table-format=1.4]
    S[table-format=1.4]}
\toprule
 &  & {DI on acceptance} & {DI} & {MSE} \\
\midrule
{Lender} & {Unconstrained} & 0.0455 & 0.1013 & 0.0934 \\
 & {Agnostic} & 0.0087 & 0.0456 & 0.1025 \\
 & {Targeted} & 0.0004 & 0.0398 & 0.1031 \\
 & {Simple} & 0.0179 & 0.0620 & 0.1103 \\
\midrule
{Regulator} & {Unconstrained} & -0.0233 & 0.0092 & 0.1077 \\
 & {Simple} & 0.0100 & 0.0348 & 0.1142 \\
\bottomrule
\end{tabular}

        }
        \caption{DI application}
        \label{tbl:results_di}
    \end{subtable}
    
    \vspace{.5cm}
    
    \begin{subtable}{\textwidth}
        \centering
                {\scriptsize	
                \addtolength{\tabcolsep}{-0.2em}
        \begin{tabular}{l
    l
    S[table-format=1.4]
    S[table-format=1.4]
    S[table-format=1.4]
    S[table-format=1.4]
    S[table-format=1.4]
    S[table-format=1.4]}
\toprule
 &  & \multicolumn{3}{c}{{High state}} & \multicolumn{3}{c}{{Low state}} \\
\cmidrule(lr){3-5} \cmidrule(lr){6-8}
 &  & {Accepted defaults} & {Accepted non-defaults} & {MSE} & {Accepted defaults} & {Accepted non-defaults} & {MSE} \\
\midrule
{Lender} & {Unconstrained} & 0.5337 & 0.8969 & 0.0932 & 0.6480 & 0.9313 & 0.1778 \\
 & {Agnostic} & 0.5510 & 0.8943 & 0.1231 & 0.6404 & 0.9343 & 0.1564 \\
 & {Targeted} & 0.5582 & 0.8932 & 0.1246 & 0.6345 & 0.9367 & 0.1555 \\
 & {Simple} & 0.6316 & 0.8824 & 0.1044 & 0.6775 & 0.9194 & 0.1959 \\
\midrule
{Regulator} & {Unconstrained} & 0.5836 & 0.8895 & 0.1305 & 0.6211 & 0.9421 & 0.1513 \\
 & {Simple} & 0.6314 & 0.8824 & 0.1333 & 0.6774 & 0.9195 & 0.1752 \\
\bottomrule
\end{tabular}
        }
        \caption{Risk application}
        \label{tbl:results_risk}
    \end{subtable}

    \vspace{.5cm}
    
    \caption{
        Performance of unconstrained and constrained prediction models across key metrics for the disparate impact (top) and risk (bottom) applications, as in \autoref{fig:results}.
        For both lender and regulator, unconstrained XGBoost and simple (five-variable) linear models are listed.
        For the lender, the agnostic model optimizes the lender objective subject to a constraint on the coefficients of the agnostic explainer, while the targeted model optimizes the objective subject to a constraint based on the targeted explainer.
        In the disparate impact application, ``DI'' denotes the average difference in credit scores across groups and ``DI on acceptance'' denotes the average in acceptance probabilities at an approval threshold of 85\% probability of repayment.
        In the risk application, ``accepted defaults'' and ``accepted non-defaults'' denotes the fraction of applicants that are accepted among those who ultimately default and among those who ultimately repay, respectively, for an approval threshold of 85\% probability of repayment.
        ``MSE'' denotes the mean-squared error.
    }
    \label{tbl:results}

\end{table}

\subsection{Optimal explainers}

\autoref{tbl:explainer} further illustrates the differences between the agnostic explainer and the optimal, targeted explainer. Recall that the agnostic explainer selects the variables that best summarize the predictions made by the lender's risk model and then projects model predictions onto these variables. The first two blocks of the table show how the regulator places constraints on the coefficients on each of the explainer variables in this projection. Within each block, the first column shows the case where the regulator constrains these coefficients, and the second column shows what value these coefficients would have taken in the lender's preferred model without any explainer constraint.

The pattern across the first two columns is illustrative. When the constrained coefficient is less positive (or more negative) than the unconstrained coefficient, the constraint has the effect of requiring the model's predicted default rates to be lower for borrowers with higher values of that variable. For example, in the first row of panel (a), the constrained coefficient on percentage of trades ever delinquent is .143, whereas the unconstrained coefficient is .256; the constraint thus lowers the credit score's predicted default rates for individuals with, for example, a 10-percentage-point higher prevalence of delinquency across their past loans by 1.1 percentage points. 

The final four columns of the table then shed light on the regulator's preferred changes in these coefficients by showing how these variables are correlated both with default (the outcome the lender's model seeks to predict) and with preference misalignment (an indicator for which borrowers the lender and regulator have divergent preferences over). Univariate correlations are shown in columns (5) and (6), while coefficients from a multivariate regression are shown in columns (7) and (8).\footnote{\label{fn:degenerate}
    Note that the last two columns show coefficients from a multivariate regression that is jointly run for all explainer covariates included in the agnostic explainer or in the targeted explainer. Due to high correlations between covariates that are only included in one but not both, some of the coefficients from the multivariate regression are degenerate. We therefore focus mainly on univariate correlations.
}

Focusing first on the disparate impact example in panel (a), where the regulator prefers \textit{lower} predicted default rates for minority applicants in order to shrink the penalty term in \eqref{eq:DI-P}, these correlations reveal that the regulator indeed tends to constrain explainer coefficients to be lower when the corresponding variable has a positive multivariate correlation with an indicator for preference misalignment (i.e., an indicator for minority status). Turning next to panel (b), where the regulator prefers \textit{higher} predicted default rates for borrowers more likely to default in a bad state of the economy, the opposite pattern is also intuitive: the regulator tends to constrain explainer coefficients to be higher when the corresponding variable is positively correlated with preference misalignment. 

Continuing to the optimal, targeted explainer, \autoref{tbl:explainer} also illustrates how the targeted explainer achieves dominant outcomes relative to the agnostic explainer. Recall that the targeted explainer selects variables that best predict preference misalignment, rather than those that best summarize model predictions, and the explainer then projects model predictions onto these variables. Similar to the agnostic case, the regulator then places constraints on the coefficients in this projection. While several variables (those in the middle of each panel) are selected by both explainers, the variables selected by only the targeted explainer tend to be more highly correlated with misalignment than the variables selected by the agnostic explainer. 
In so doing, the targeted explainer leads to model predictions that are more closely aligned with the regulator's preferences, at lower cost in terms of overall model performance, relative to an agnostic explainer. 

There are further interesting patterns in which variables are selected by the agnostic and targeted explainers. In panel (a), the disparate impact case, we see that the main drivers of differences between groups and the main drivers of repayment are quite different, such that the agnostic and targeted explainers describe substantially different aspects of the credit scoring function. In particular, the agnostic explainer selects variables that capture a history of credit risk and hence may be particularly predictive of future credit risk; in contrast, the targeted explainer selects variables that reflect broader patterns in credit use and access, which may correlate more closely with the drivers of inequality (e.g., historical exclusion from formal credit markets) that shape the regulator's group-specific preferences. In the risk case (panel (b)), on the other hand, the targeted explainer picks variables that are particularly helpful to describe who potentially defaults in the low state of the economy. These variables are more similar to the variables selected by the agnostic explainer in the sense that both capture a history of credit risk and are selected to be predictive of future risk. These differences help interpret why the choice of explainer makes a relatively minor difference in the risk case, as in Figure \ref{fig:results}.

\begin{landscape}
\begin{table}[h]
\centering
    \begin{subtable}{1.4\textwidth}
        \centering
        \resizebox{\textwidth}{!}{\begin{tabular}{>{\hangindent=2em}p{36em}
    S[table-format=1.4]
    S[table-format=1.4]
    S[table-format=1.4]
    S[table-format=1.4]
    S[table-format=1.4]
    S[table-format=1.4]
    S[table-format=1.4]
    S[table-format=1.4]}
\toprule
 & \multicolumn{2}{c}{{Agnostic explainer}} & \multicolumn{2}{c}{{Targeted explainer}} & \multicolumn{2}{c}{{Correlation}} & \multicolumn{2}{c}{{Coefficients}} \\
\cmidrule(lr){2-3} \cmidrule(lr){4-5} \cmidrule(lr){6-7} \cmidrule(lr){8-9}
 & {Constrained} & {Unconstrained} & {Constrained} & {Unconstrained} & {Default} & {Misalignment} & {Default} & {Misalignment} \\
\midrule
Intercept & 0.1511 & 0.1525 & 0.1511 & 0.1525 &  &  & 0.1501 & 0.1973 \\
\midrule
Percentage of trades ever delinquent & 0.1433 & 0.2563 &  &  & 0.3569 & 0.1719 & 0.0432 & 0.0189 \\
Missing: Months since most recent third party collection occurrence & -0.0111 & -0.0334 &  &  & -0.3097 & -0.1769 & -0.0276 & -0.0066 \\
Number of trades 60 or more days past due ever & -0.1300 & -0.2704 &  &  & 0.3346 & 0.1218 & 0.0504 & -0.0079 \\
Missing: Balance of most recent, open, premium bankcard trade in 12 months & -0.0026 & 0.0542 & 0.0014 & -0.0082 & 0.2618 & 0.2062 & 9.5044 & -89.1644 \\
Missing: Months since most recent non-medical third party collection occurrence & -0.0196 & -0.0470 & -0.0474 & -0.0965 & -0.3073 & -0.1812 & -0.0247 & -0.0310 \\
Missing: Credit line of most recent, open, premium bankcard trade in 12 Months &  &  & 0.0014 & -0.0082 & 0.2618 & 0.2062 & -9.4854 & 89.1995 \\
Average bankcard credit limit &  &  & 0.0079 & -0.0398 & -0.1955 & -0.1803 & -0.0115 & -0.0289 \\
Transactor behavior in past 8 months &  &  & -0.0104 & -0.0500 & -0.2125 & -0.1674 & -0.0227 & -0.0291 \\
\bottomrule
\end{tabular}}
        \caption{DI application}
        \label{tbl:explainer_di}
    \end{subtable}
    
    \vspace{.5cm}
    
    \begin{subtable}{1.4\textwidth}
        \centering
        \resizebox{\textwidth}{!}{\begin{tabular}{>{\hangindent=2em}p{36em}
    S[table-format=1.4]
    S[table-format=1.4]
    S[table-format=1.4]
    S[table-format=1.4]
    S[table-format=1.4]
    S[table-format=1.4]
    S[table-format=1.4]
    S[table-format=1.4]}
\toprule
 & \multicolumn{2}{c}{{Agnostic explainer}} & \multicolumn{2}{c}{{Targeted explainer}} & \multicolumn{2}{c}{{Correlation}} & \multicolumn{2}{c}{{Coefficients}} \\
\cmidrule(lr){2-3} \cmidrule(lr){4-5} \cmidrule(lr){6-7} \cmidrule(lr){8-9}
 & {Constrained} & {Unconstrained} & {Constrained} & {Unconstrained} & {Default} & {Misalignment} & {Default} & {Misalignment} \\
\midrule
Intercept & 0.2608 & 0.1153 & 0.2683 & 0.1499 &  &  & 0.1503 & 0.1458 \\
\midrule
Missing: Balance of Most Recent, Open, Premium Bankcard Trade in 12 Months & 0.0792 & 0.0499 &  &  & 0.2848 & 0.1478 & 0.0349 & 0.0091 \\
Missing: Months since most recent third party collection occurrence & -0.0542 & -0.0263 &  &  & -0.3287 & -0.1941 & -0.0247 & -0.0155 \\
Missing: Months since most recent non-medical third party collection occurrence & -0.0535 & -0.0269 &  &  & -0.3268 & -0.1887 & -0.0193 & -0.0021 \\
Number of trades 60 or more days past due ever & -0.3562 & -0.1813 &  &  & 0.3467 & 0.1085 & 0.0535 & -0.0140 \\
Percentage of trades ever delinquent & 0.2824 & 0.1472 & 0.1503 & 0.0964 & 0.3749 & 0.2058 & 0.0387 & 0.0315 \\
Missing: Max aggregate bankcard balance over last 3 months &  &  & 0.1471 & 0.0688 & 0.1907 & 0.2040 & 0.0074 & 0.0152 \\
Number of unpaid collections &  &  & 0.0467 & 0.0203 & 0.2767 & 0.2047 & 0.0271 & 0.0265 \\
Missing: Bankcard balance magnitude algorithm over last 24 months &  &  & 0.0536 & 0.0115 & 0.1722 & 0.1833 & 0.0121 & 0.0123 \\
Missing: Months since most recent bankcard trade opened &  &  & -0.2209 & -0.0919 & 0.2146 & 0.2101 & 0.0008 & 0.0191 \\
\bottomrule
\end{tabular}}
        \caption{Risk application}
        \label{tbl:explainer_risk}
    \end{subtable}
    
    \vspace{.5cm}
    
    \caption{
        Explainer coefficients in the disparate impact (top) and risk (bottom) applicants for the solutions presented in \autoref{fig:results} and \autoref{tbl:results}.
        The ``agnostic explainer'' and ``targeted explainer'' columns report the coefficients on the respective five chosen explainer variables and a constant for different credit scores.
        The ``constrained'' coefficients report the coefficients for the regulator's preferred model, which are also those of the lender's model subject to the explanation constraint (dot at the end of the blue and green lines, respectively, in \autoref{fig:results}).
        The ``unconstrained'' coefficients report the explainer coefficients for the unconstrained lender model (lender dot at the end of the solid black lines in \autoref{fig:results}).
        The remaining columns provide details of the chosen variables, listing the correlation with default and the variable encoding misalignment (that is, minority identity in the disparate impact application and a rejected application in the risk application), as well as linear regression coefficients (discussed further in \hyperref[fn:degenerate]{Footnote~\ref{fn:degenerate}} of the text)  for a joint linear regression of the variable on all covariates in the table.
    }
    \label{tbl:explainer}
\end{table}
\end{landscape}

\FloatBarrier

\section{Conclusion}
\label{sec:conclusion}

We study the problem of a principal who seeks to regulate an agent's choices over complex algorithms. In an empirical example, we consider a lender (agent) who is choosing a complex credit scoring function to evaluate the credit risk of potential borrowers, whereas a financial regulator (principal) seeks either to reduce disparate impact of the credit scoring function across racial and ethnic groups or to make the credit scoring function more conservative over a group of borrowers who may be particularly likely to default in a bad state of the economy. The example illustrates our theoretical results:
restricting the lender ex-ante to simple credit-scoring algorithms makes regulation easier, but creates inefficient results that can come at a cost to both the regulator and the lender.
Instead, we show that effective regulation can allow for complex machine-learning algorithms, and should regulate them based on those aspects of the model most related to the source of preference misalignment.
We thus add to a broader discussion about regulating algorithms:
when should regulators allow the use of complex models, and what aspects of these models should regulators regulate? 
Our theory and empirical results add further evidence that regulation based on prohibiting the use of certain data or constraining the functional form of models may be ineffective and inefficient.
Instead, our work illustrates benefits to allowing for flexible algorithms that are regulated based on descriptions of their key behaviors, and our results provide guidance for what these key behaviors are.

\bigskip

--

\textsc{Booth School of Business, University of Chicago}, Chicago, IL, USA 

\textsc{Graduate School of Business, Stanford University}, Stanford, CA, USA

\clearpage
\appendix
\section*{Appendix}

\renewcommand{\thesubsection}{\Alph{subsection}}

\subsection{Empirical optimization}
\label{apx:optimization}

In empirically solving for optimal credit scores, we face three challenges.
First, we only observe a sample, and not the full distribution of the data.
Second, solving the regulator's optimization problem of finding good explainer covariates is generally computationally hard.
And third, we have to solve for complex prediction functions subject to explainer constraints.

\begin{figure}[h]
    \centering

    \begin{subfigure}[b]{\textwidth} %
        \includegraphics[width=\textwidth]{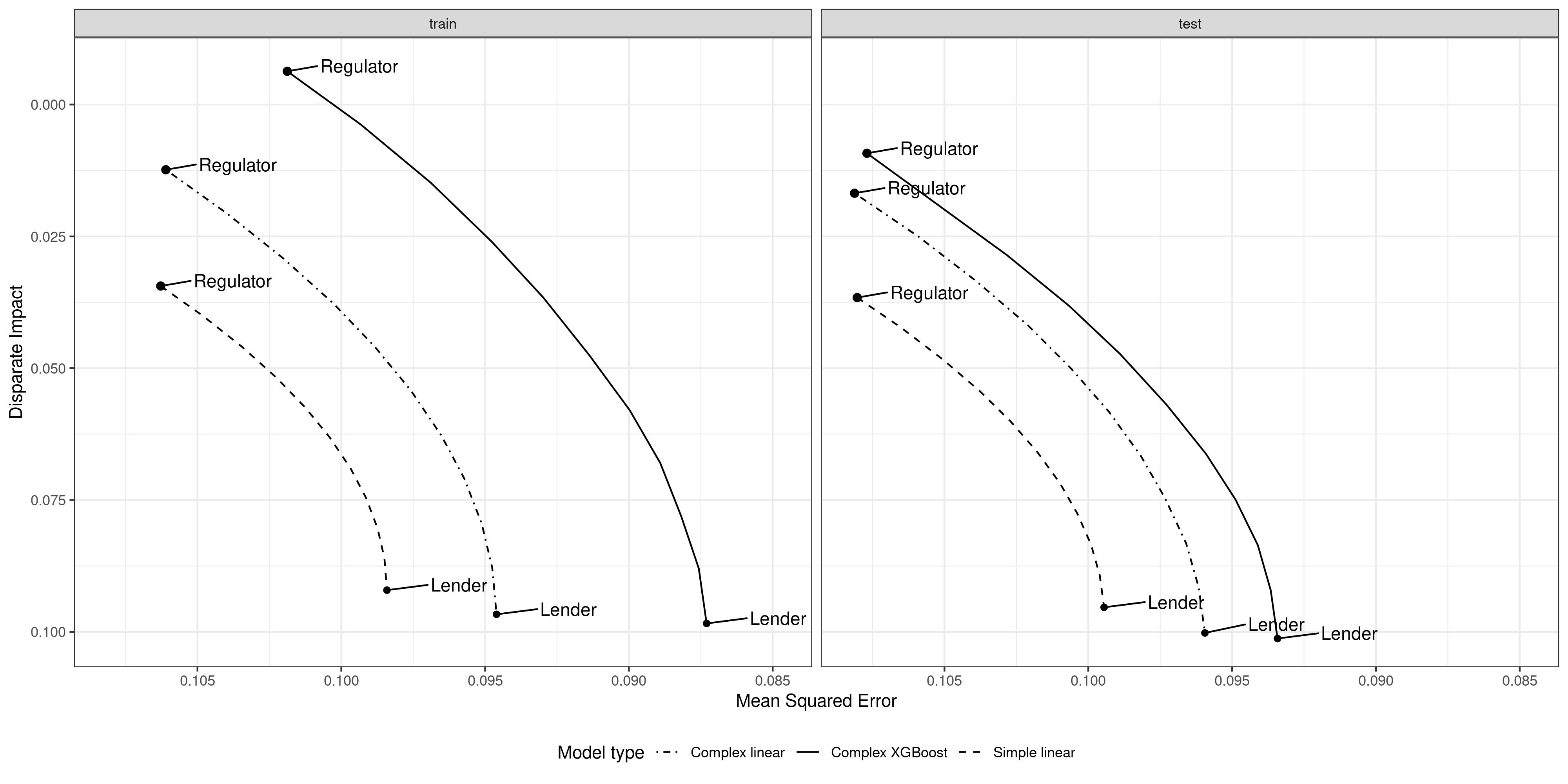}
        \caption{DI application}
        \label{fig:traintest_di}
    \end{subfigure}
    
    \vspace{.5cm}

    \begin{subfigure}[b]{\textwidth} %
        \includegraphics[width=\textwidth]{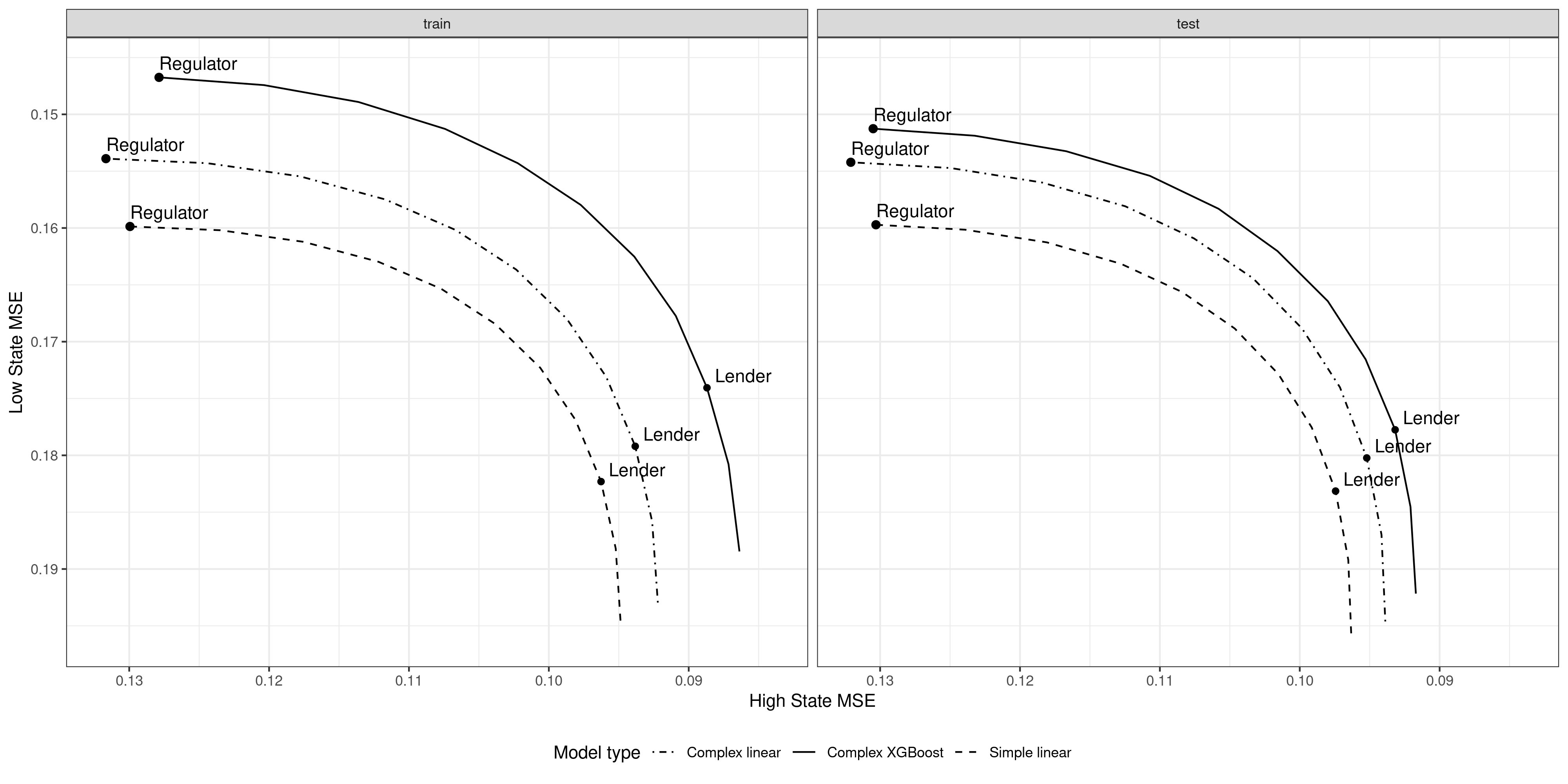}
        \caption{Risk application}
        \label{fig:traintest_risk}
    \end{subfigure}

    \vspace{.5cm}
    
    \caption{
        Performance of models of varying complexity in the training sample (left panels) and on the hold-out set (right panels) for the disparate impact (top panels) and risk (bottom panels) examples, following \autoref{fig:results}.
        The simple linear model is a linear-regression model on five covariates chosen by the LASSO.
        The complex linear and complex XGBoost models use all 518 covariates to predict repayment.
        The regulator and lender marks correspond to the solutions maximizing the empirical analog to the objectives \eqref{eq:DI-P}, \eqref{eq:DI-A} and \eqref{eq:R-P}, \eqref{eq:R-A}, respectively.}
    \label{fig:traintest}
\end{figure}

In order to address the first challenge of facing only a noisy sample form the distribution of applicants, we solve all of our optimization problems in a training sample, and later evaluate the solutions on a hold-out set to obtain accurate measurements of performance.
We implement the above solutions of the lender and regulator using boosted trees, which has the best out-of-sample performance among a number of parametric and non-parametric standard prediction methods on our data.
We provide additional details in the appendix, including a comparison to regularized and non-regularized linear regression, random forests, and neural networks in \autoref{tbl:performance} and a comparison of in-sample and out-of-sample performance of our solutions in \autoref{fig:traintest}.
In the remaining part of this section, all reported results are based on the out-of-sample performance of boosted trees.

For the second challenge, obtaining the optimal set of explainer variables in each case involves a search over all possible sets of $k$ variables, where $k=5$ in our main specification, out of a total of more than 500.
This would generally be computationally infeasible.
Instead, we choose the set of variables by a LASSO regression that approximates the solution of \eqref{eq:Ex}, \eqref{eq:R-Ex}, and \eqref{eq:DI-Ex}.
Specifically, we fit linear models that minimize the sum of the respective empirical risk and a penalty $\nu \: \|\beta\|_1$ on the coefficients $\beta$, where $\nu$ is chosen such that the number of non-zero coefficients is $k$. We then use the set of variables with non-zero coefficients as the explainer variables.

To address the third challenge of solving for complex prediction functions subject to explainer constraints, we rely on projection methods.
We first solve for the first-best lender and regulator solutions using boosted trees (XGBoost) on the training sample, as well as for preferences that put varying weight $\lambda$ on disparate impact in the DI application and preferences that put varying probability $\ell$ on the low-state MSE in the risk application.
We then compute the respective solution subject to explainer constraints by applying \autoref{prop:constrained}.
Specifically, from an unconstrained solution ${f}$ we obtain the constrained solution by setting $\hat{f}(x) = x_S'\hat{\beta}^P + ({f}(x) - x_S'\hat{\beta}^{{f}})$, where $\hat{\beta}^P$ are the OLS coefficients for the regression of the first-best regulator solution on the explainer covariates in $S$ and $\hat{\beta}^{{f}}$ are the analogous coefficients for ${f}$. 

\clearpage

\subsection{Additional Empirical Results}

\begin{table}[h]
    \centering
    \small
    \begin{tabular}{lccc}
\toprule
 & {Accepted} & {Rejected} & {Minority} \\
\midrule
Repayment & 0.849 & -- & 0.841 \\
 & (0.358) &  & (0.431) \\
\midrule
Percentage of trades ever delinquent & 12.005 & 24.236 & 18.890 \\
 & (19.706) & (28.382) & (23.951) \\
Missing: Months since most recent third party collection occurrence & 0.752 & 0.510 & 0.596 \\
 & (0.432) & (0.500) & (0.491) \\
Missing: Months since most recent non-medical third party collection occurrence & 0.823 & 0.612 & 0.683 \\
 & (0.381) & (0.487) & (0.465) \\
At least one trade 90 or more days past due ever & 0.269 & 0.478 & 0.424 \\
 & (0.443) & (0.500) & (0.494) \\
Missing: Aggregate bankcard amount past due for month 17 & 0.190 & 0.374 & 0.306 \\
 & (0.392) & (0.484) & (0.461) \\
Missing: Aggregate bankcard amount past due for month 15 & 0.187 & 0.371 & 0.303 \\
 & (0.390) & (0.483) & (0.460) \\
Number of collections excluding medical & 0.529 & 1.446 & 1.032 \\
 & (1.575) & (2.727) & (2.168) \\
Missing: Aggregate bankcard amount past due for month 18 & 0.190 & 0.372 & 0.306 \\
 & (0.393) & (0.483) & (0.461) \\
Number of unpaid collections & 0.715 & 2.384 & 1.313 \\
 & (2.326) & (4.978) & (3.015) \\
Missing: Number of months since overlimit on a bankcard & 0.760 & 0.673 & 0.696 \\
 & (0.427) & (0.469) & (0.460) \\
\bottomrule
\end{tabular}
    \caption{Summary statistics of repayment as well as the ten variables (out of a total of 518) with the highest importance scores for predicting credit-card repayment. Averages are reported separately for subsamples that were accepted or rejected for a loan. An additional column presents average values specifically for the subsample of accepted minority applicants. Standard deviations are in parentheses.
    }
    \label{tbl:descriptives}
\end{table}

\begin{table}[h]
    \centering
    \begin{tabular}{l
       S[table-format=1.4]
       S[table-format=1.4]
       S[table-format=1.4]
       S[table-format=1.4]}
\toprule
Model & {MSE} & {ROC AUC} & {Log loss} & {$R^2$} \\
\midrule
{XGBoost} & 0.0934 & 0.8672 & 0.3007 & 0.2802 \\
\midrule
{Logistic XGBoost} & 0.0935 & 0.8680 & 0.3009 & 0.2794 \\
{Linear model} & 0.0959 & 0.8603 & 0.3230 & 0.2607 \\
{Elastic net} & 0.0959 & 0.8609 & 0.3210 & 0.2610 \\
{Logistic model} & 0.0955 & 0.8626 & 0.3103 & 0.2639 \\
{Random forest} & 0.0946 & 0.8608 & 0.3092 & 0.2711 \\
{Logistic random forest} & 0.0945 & 0.8632 & 0.3057 & 0.2715 \\
{Neural network} & 0.0973 & 0.8504 & 0.3569 & 0.2501 \\
{Logistic neural network} & 0.0950 & 0.8637 & 0.3045 & 0.2683 \\
\midrule
{Finscore logistic} & 0.1031 & 0.8429 & 0.3354 & 0.2058 \\
\bottomrule
\end{tabular}
    \caption{Out-of-sample performance of varying prediction models according to mean-squared error (``MSE,'' where lower values indicate better fit), the area under the ROC curve (``ROC AUC,'' where higher values indicate better fit), the negative log-likelihood (``log loss,'' where lower values indicate better fit), and the coefficient of determination (``$R^2$,'' where higher values indicate better fit). Our main implementation is based on boosted trees (``XGBoost,'' first row).
    As a benchmark, we also report results for using a commercially available credit score as the only covariate in a simple logistic regression (``finscore logistic,'' last row). This credit score is not used as a covariate in the other models.
    }
    \label{tbl:performance}
\end{table}

\FloatBarrier

\begin{figure}[h]
    \centering
    
    \begin{subfigure}[b]{\textwidth}
    \centering
    \begin{subfigure}[b]{0.45\textwidth}
        \centering
        \includegraphics[width=\textwidth]{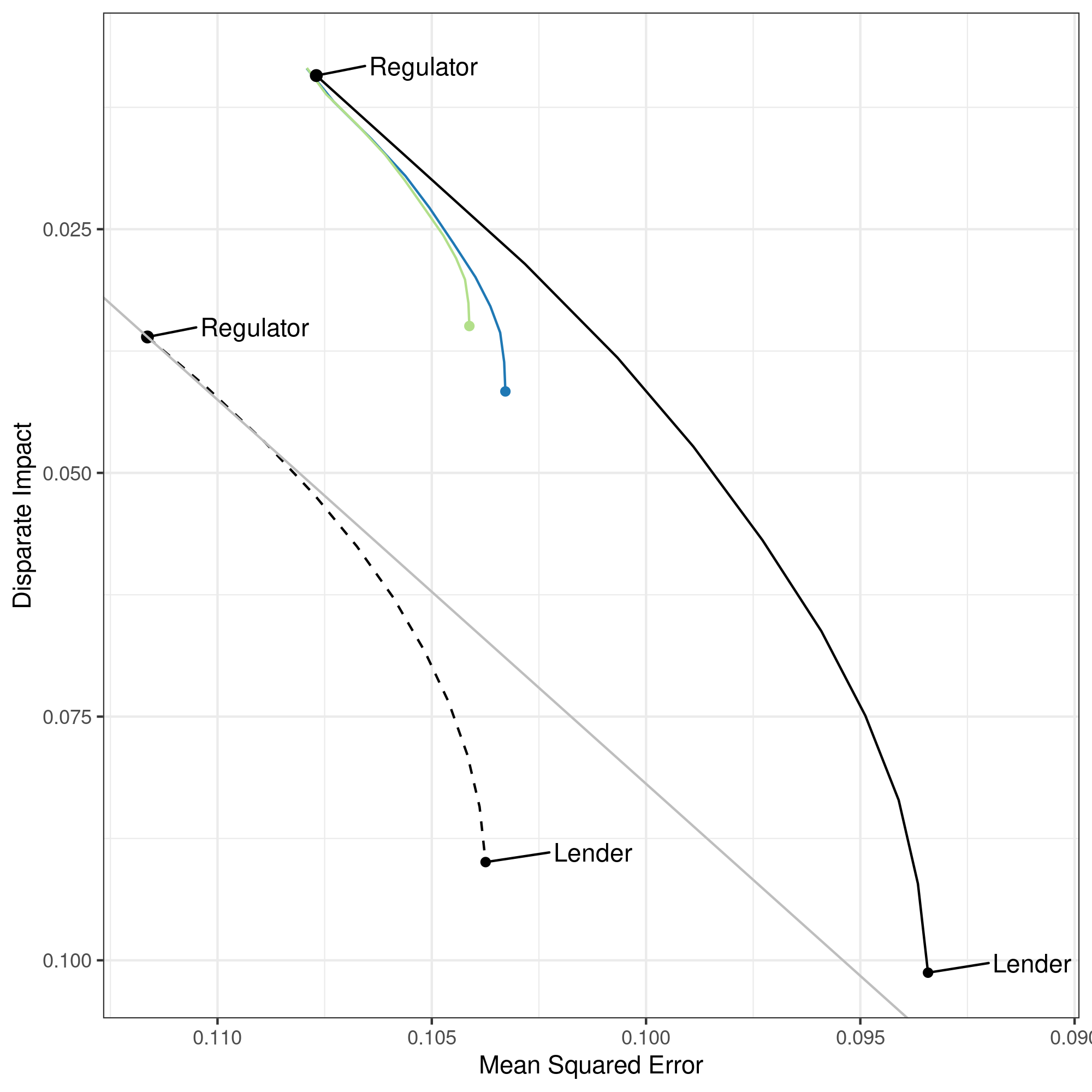}
        \caption*{Ten explainer variables}
    \end{subfigure}
    \hfill
    \begin{subfigure}[b]{0.45\textwidth}
        \centering
        \includegraphics[width=\textwidth]{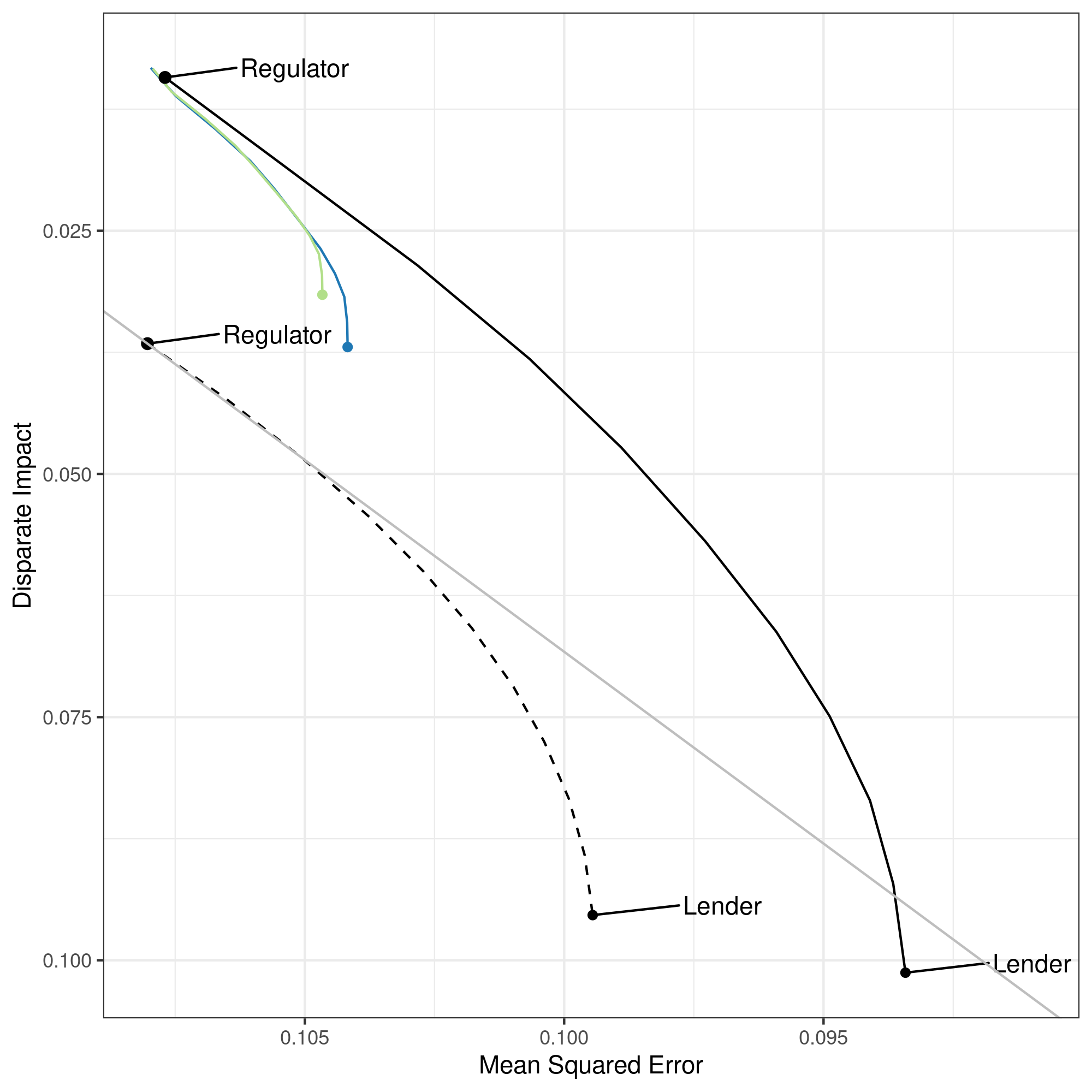}   %
        \caption*{Twenty explainer variables}
    \end{subfigure}
    
    \caption{DI application}
    \label{fig:complexity_di}
    \end{subfigure}
    
    \vspace{0.5cm}  %
    
    \begin{subfigure}[b]{\textwidth}
    \centering
    \begin{subfigure}[b]{0.45\textwidth}
        \centering
        \includegraphics[width=\textwidth]{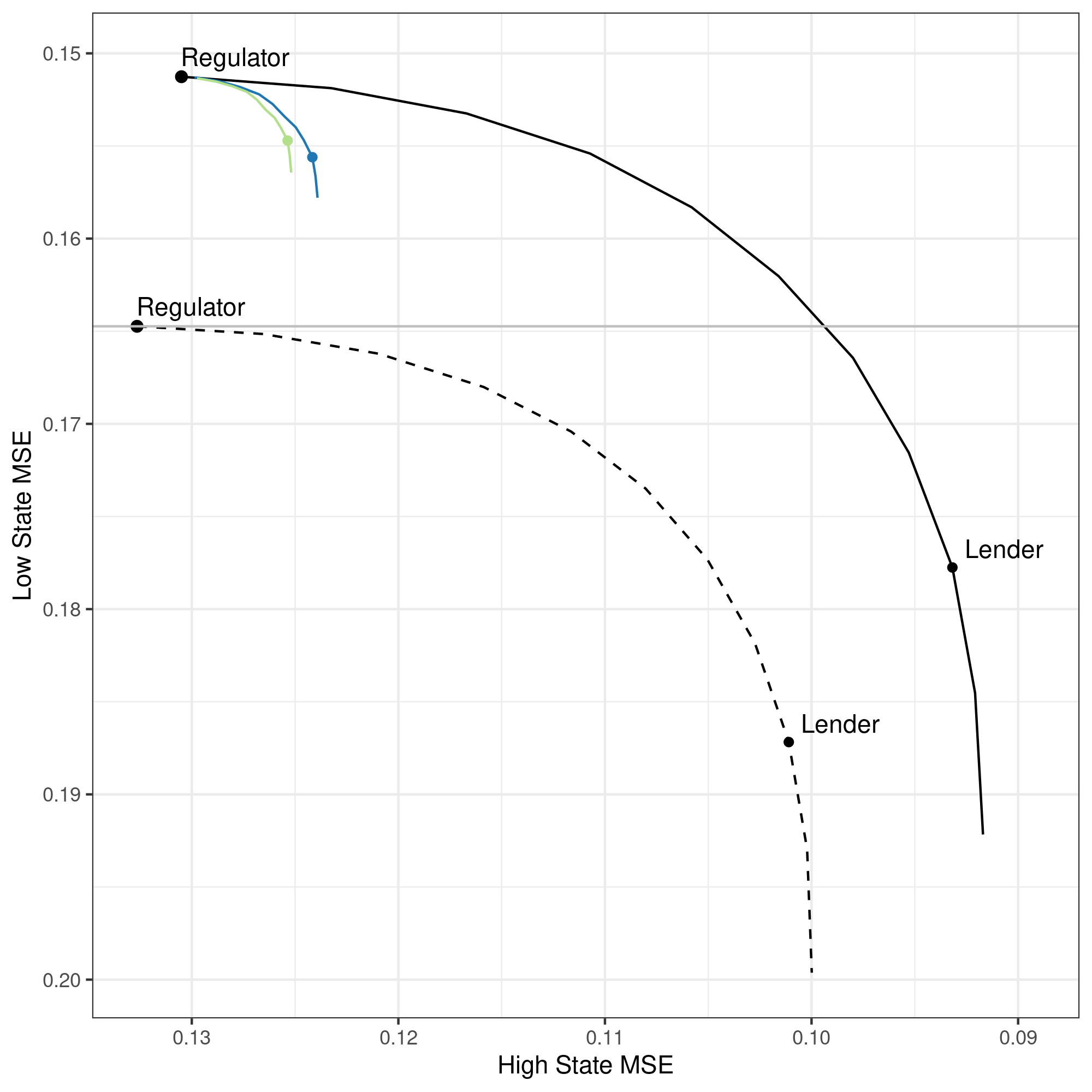}
        \caption*{Ten explainer variables}
    \end{subfigure}
    \hfill
    \begin{subfigure}[b]{0.45\textwidth}
        \centering
        \includegraphics[width=\textwidth]{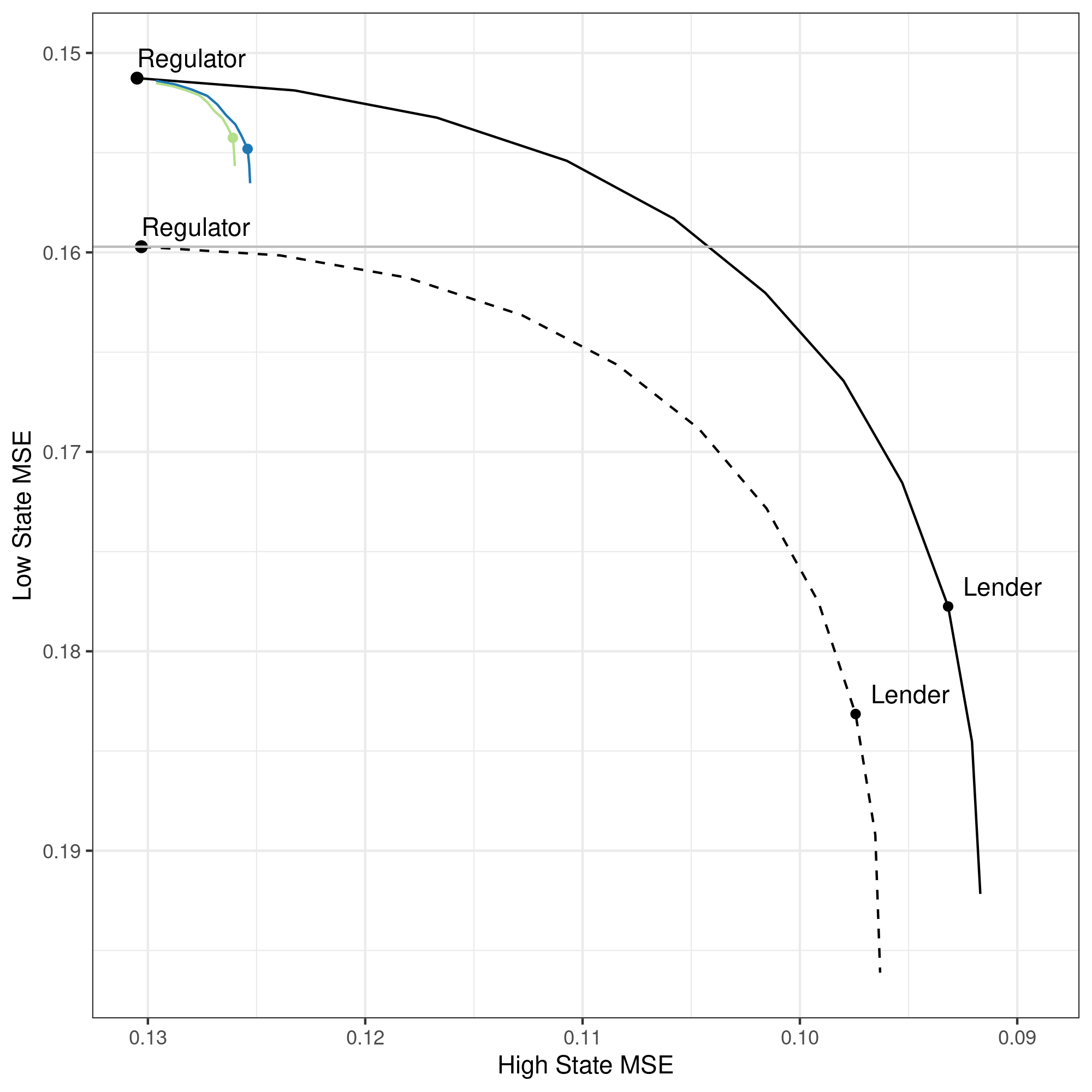}   %
        \caption*{Twenty explainer variables}
    \end{subfigure}
    
    \caption{Risk application}
    \label{fig:complexity_risk}
    \end{subfigure}
    
    \vspace{0.5cm}

        \includegraphics[width=\textwidth]{graphs/legend.pdf}
    
    \caption{Out-of-sample performance of unconstrained and constrained prediction models across key objectives for the disparate impact (top) and risk (bottom) applications as in \autoref{fig:results}, but with varying explainer complexity.
    The left panels use ten variables for the simple (linear) model and the explainer (in addition to a constant for the intercept), while the right panel uses twenty.
    This illustrates robustness of the results reported in \autoref{tbl:results} and \autoref{fig:results} based on five explainer variables.}

    \label{fig:complexity}
\end{figure}

\FloatBarrier

\clearpage

\subsection{Proofs}

\begin{proof}[Proof of \autoref{prop:constrained}]
    Write $s(x) = x'_S \E_\theta^{-1}[x_S x_S'] \E_\theta[x_S y]$ for the linear projection of $y$ on $x_S$.
    Among the constrained set, the agent chooses among functions $x_S'\beta_\theta^P + r(x)$ subject to $\E_\theta[r(x) x_S] = \0$, minimizing
    \begin{align*}
        \E_\theta[(x_S'\beta_\theta^P + r(x) - y)^2]
        &=
        \E_\theta[(x_S'\beta_\theta^P - s(x) + r(x) - (y - s(x)))^2]
        \\
        &=
        \E_\theta[(x_S'\beta_\theta^P - s(x))^2
        + \E_\theta[(r(x) - (y - s(x)))^2].
    \end{align*}
    Hence, the optimality of $r(x)$ does not depend on $\beta_\theta^P$, so $r(x) = r_\theta^A(x)$ is an optimal solution.
\end{proof}

\begin{proof}[Proof of \autoref{thm:optimalregulation}]
    For \emph{covariate shift}, $\F = \R^\X$ achieves the first-best solution $f_\theta$ on the support of $\mu_\theta$, and thus also for the principal.
    Only if the ex-ante complexity constraint is trivial in the sense that it allows for solutions that are $\mu_\theta^P$-almost surely the same as $f_\theta$ would a constrained solution achieve the same utility.
    For \emph{model shift}, by \autoref{prop:constrained}, the solution from delegating to the agent subject to an explainability constraint with covariates $\Sexplain$ takes the form $f_\theta(x) - x_{\Sexplain}'\beta$ for some $\beta$ that the principal has control over by requiring $\E f = \ex f_\theta - \beta$.
    Hence, the principal achieves risk $\int_\X (f_\theta(x) - x_{\Sexplain}'\beta - f_\theta^P(x)) \d \mu_\theta(x)$, where $\beta$ can be set to attain the minimum, yielding the left-hand risk for the principal.
    At the same time, if the principal restricts the agent to simple linear functions on covariates $\Srestrict$, she can dictate the choice of prediction function and achieve risk $\int_\X (x_{\Srestrict}'\beta - f_\theta^P(x)) \d \mu_\theta(x)$, where $\beta$ can be set to attain the minimum again, yielding the right-hand risk for the principal.
\end{proof}

\begin{proof}[Proof of \autoref{prop:disparate}]
    The preference of the principal is equivalent to minimizing
    \begin{align*}
        &\int_X (f(x) - f_\theta(x))^2 \d \mu_\theta(x)
        + \lambda \left(\int_X f(x) \frac{g_\theta(x)}{\bar{g}_\theta} \d \mu_\theta(x) - \int_\X f(x) \frac{1 - g_\theta(x)}{1 - \bar{g}_\theta} \d \mu_\theta(x) \right)
        \\
        &=
        \int_X \left((f(x) - f_\theta(x))^2
        + \lambda f(x) \left(\frac{g_\theta(x)}{\bar{g}_\theta} - \frac{1 - g_\theta(x)}{1 - \bar{g}_\theta}\right)\right)\d \mu_\theta(x)
        \\
        &=
        \int_X \left(f^2(x) - 2 f(x) 
        \left(
        f_\theta(x)
        - \lambda\frac{g_\theta(x) - \bar{g}_\theta}{2 \bar{g}_\theta (1-\bar{g}_\theta)} 
        \right)
        + f_\theta^2(x)\right)
        \d \mu_\theta(x)
        \\
        &=
        \int_X
        \left(
        f(x)
        -
        f_\theta(x)
        + \lambda\frac{g_\theta(x) - \bar{g}_\theta}{2 \bar{g}_\theta (1-\bar{g}_\theta)} 
        \right)^2
        \d \mu_\theta(x)
        +
        \int_X \left( 
        f_\theta^2(x)
        -
        \left(
        f_\theta(x)
        - \lambda\frac{g_\theta(x) - \bar{g}_\theta}{2 \bar{g}_\theta (1-\bar{g}_\theta)} 
        \right)^2
        \right)
        \d \mu_\theta(x).
    \end{align*}
    Since the right part does not depend on $f_\theta$, this preference is equivalent to \autoref{def:modelshift} with $f_\theta^P(x) = f_\theta(x)
    - \lambda\frac{g_\theta(x) - \bar{g}_\theta}{2 \bar{g}_\theta (1-\bar{g}_\theta)}$.
\end{proof}

\begin{proof}[Proof of \autoref{thm:optimalexplainer}]
    The result is immediate from the fact that the expression the theorem maximizes over represents the average risk from optimal delegation with explainer, as in the proof of \autoref{thm:optimalregulation}.
\end{proof}

\clearpage
\bibliography{literature}

\end{document}